\documentclass[notitlepage]{article}

\usepackage{amsfonts}
\usepackage{amsmath}
\usepackage{units}
\usepackage{amssymb}
\usepackage{graphicx}
\usepackage{epstopdf}
\usepackage[top=1.5in, bottom=1in, left=1in, right=1in]{geometry}

\usepackage{titling}

\title{Spin Dressed Relaxation and Frequency Shifts from Field Imperfections.}
\author{C.M. Swank, E.K. Webb, X. Liu and B.W. Filippone}
\begin{document}
\begin{titlingpage}
\maketitle

\begin{abstract}
Critical dressing, the simultaneous dressing of two spin species to the same effective Larmor frequency,  is a technique that can, in principle, improve the sensitivity to small frequency shifts. The benefits of spin dressing and thus critical dressing are achieved at the expense of generating a large (relative to the holding field $B_{0}$,) homogeneous oscillating field. Due to inevitable imperfections of the fields generated, the benefits of spin dressing may be lost from the additional relaxation and noise generated by the dressing field imperfections. In this analysis the subject of relaxation and frequency shifts are approached with simulations and theory. Analytical predictions are made from a new quasi-quantum model that includes gradients in the holding field $B_{0}=\omega _{0}/\gamma $ and dressing field $B_{1}=\omega _{1}/\gamma $ where $B_{1}$ is oscillating at frequency $\omega $. It is found that irreversible DC gradient relaxation can be canceled by an AC spin dressing gradient in the Redfield regime. Furthermore, it is shown that there is no linear in $E$ frequency shift generated by gradients in the dressing field. The results are compared with a Monte Carlo simulation coupled with a 5$^{\text{th}}$ order Runge-Kutta integrator. Comparisons of the two methods are presented as well as a set of optimized parameters that produce stable critical dressing at a range for oscillating frequencies $\omega ,$ as well as pulsed frequency modulation parameters for maximum sensitivity.
\end{abstract}
\tableofcontents
\end{titlingpage}

\section{Introduction}

Spin Dressing is a technique that changes the effective Zeeman splitting of an atomic or nuclear spin system  \cite{CohenBook,CohenHarocheDressing1969}, and is found to have a wide range of applications. Notably, the manipulation of spin dynamics of ultracold quantum gases \cite{Gerbier2006}, and orbit interactions and its influence on dynamics, and enabling a new avenue for coupling optics to nano-materials \cite{Pervishko2015}. It is also a valuable tool in the field of fundamental physics. 

A variation of spin dressing has been proposed as a technique to maximize sensitivity in the search for the time reversal and parity violating observable, the permanent electric dipole moment (EDM). In reference \cite{golub1994} they propose critical dressing of spin-polarized neutrons and $^3$He. Critical dressing is the simultaneous spin dressing of two species so that they have the same effective gyromagnetic ratio. 

Here we present a detailed framework of the technique of critical dressing with the aim of optimization of experimental sensitivity and mitigation of systematic effects. This is achieved by including the effects of the fluctuations in field as observed by the particle's trajectory through the field inhomogeneities, similar to the derivation in reference \cite{redfield}. In reference \cite{CohenCollision} they approach the case of relaxation due to collisions of gas atoms, however in the system presented here, scattering in the bulk does not directly affect the dynamics, it only modifies the field fluctuations observed by the particle by modification of the particle's trajectory.  In reference \cite{BevilacquaDressing} they reformulate the result in \cite{CohenHarocheDressing1969} and find solutions for non-harmonic dressing fields as well offset dressing fields. In reference \cite{HartemultiFdress2018}, they consider the case of a chromatic dressing field and in \cite{Yuen2018} they find the propagator for the polychromatic dressed Hamiltonian. Here we consider fluctuations in the field producing the spin dressing of a gas sample over a macroscopically large volume, where the particle's trajectory through the inhomogeneities of the DC holding field, as well as the spin dressing field, are the largest sources of relaxation and frequency shifts. We use an approach similar to that found in references \cite{redfield}, \cite{GolubSteyerlRedfield} and \cite{pignol2015}. We show that the spectrum of the correlation functions that determine the relaxation and frequency shifts depend on the dressed energy, it is not enough to correct the result in the case of no oscillating field by a factor $J_0(\frac{\omega_1}{\omega}),$ as would be found if one included the shift as an intrinsic energy splitting in the Hamiltonian.

The benefits of spin dressing and thus critical dressing are achieved at the expense of generating a large (relative to the holding field $B_{0}$,) homogeneous oscillating field. Due to inevitable imperfections of laboratory fields, the benefits of the spin dressing technique may be lost from the additional relaxation and noise generated by an imperfect field. In this
analysis the subject of relaxation and frequency shifts are approached with a quasi-quantum model that includes gradients in the holding field $B_{0}=\omega _{0}/\gamma $ and dressing field $B_{1}=\omega _{1}/\gamma $ oscillating at frequency $\omega $. We will present an analytic model that can determine the relaxation and frequency shifts given an inhomogeneity of the holding field $B_0$ and a uniform applied electric field. Furthermore we compare the results to a Monte Carlo simulation coupled with a 5$^{\text{th}}$ order Runge-Kutta integrator. This simulation is a modified version of the simulation outlined in reference \cite{riccardoThesis} and used to produce the results in reference \cite{SchmidERelax2008}. This code was modified to produce a set of parameters that produce critical dressing at a range of oscillating
frequencies $\omega ,$ as well as parameters for pulsed frequency modulation of the spin dressing field.

\section{Spin Dressing Model}

Starting from the Hamiltonian, according to references  \cite{CohenBook,CohenHarocheDressing1969}, for a spin in a strong AC magnetic field along $x$ ($B_1$) and weaker DC holding field ($B_0$) along $z$,
\begin{align}
\frac{H}{\hbar }=\omega a^{\dagger }a+\frac{\gamma B_{1}}{2\lambda ^{1/2}}%
\frac{\sigma _{x}}{2}\left( a+a^{\dagger }\right) +\omega _{0}\frac{\sigma
_{z}}{2}
\end{align}

where 
\begin{align}
\lambda =\frac{B_{1}^{2}V}{8\pi \hbar \omega }.
\end{align}
The first term represents the energy in the oscillating field, the second term represents the energy of the interaction of the oscillating field with the spin, and the third term is the energy of the interaction of the spin with the holding field. The Pauli matrices in this system are chosen so that the diagonal state is in the direction of the strong oscillating field,

\begin{align}
\sigma _{x} &=\left( 
\begin{array}{cc}
1 & 0 \\ 
0 & -1%
\end{array}%
\right) =\left\vert +\right\rangle \left\langle +\right\vert -\left\vert
-\right\rangle \left\langle -\right\vert , \\
\sigma _{y} &=\left( 
\begin{array}{cc}
0 & 1 \\ 
1 & 0%
\end{array}%
\right) =\left\vert +\right\rangle \left\langle -\right\vert +\left\vert
-\right\rangle \left\langle +\right\vert , \\
\sigma _{z} &=\left( 
\begin{array}{cc}
0 & -i \\ 
i & 0%
\end{array}%
\right) =-i\left\vert +\right\rangle \left\langle -\right\vert +i\left\vert
-\right\rangle \left\langle +\right\vert .
\end{align}

\bigskip An ensemble of photons of a harmonic field is described by Glauber states such that the coefficients of the probability amplitudes in terms of the average photon number $\,\,\,\lambda =\left\langle n\right\rangle $ are

\begin{align}
a_{n} &=\exp (-\frac{\lambda }{2})\frac{\lambda ^{\frac{n}{2}}}{\left(
n!\right) ^{\frac{1}{2}}},
\end{align}

The basis in $m_{z}$ can be written in terms of the $m_{x}~$eigen states,
\begin{align}
\overline{\left\vert n,m_{z}\right\rangle }=\frac{1}{\sqrt{2}}\overline{%
\left\vert n_{+}\right\rangle }\left\vert +\right\rangle _{x}+im_z\overline{%
\left\vert n_{-}\right\rangle }\left\vert -\right\rangle _{x}
\end{align}%
where 
\begin{align}
\overline{\left\vert n_{+}\right\rangle } &=e^{-\frac{1}{2}\frac{\eta }{%
\omega }(a^{\dagger }-a)}\left\vert n\right\rangle,\\
\overline{\left\vert n_{-}\right\rangle } &=e^{\frac{1}{2}\frac{\eta }{%
\omega }(a^{\dagger }-a)}\left\vert n\right\rangle.
\end{align}
and $\left\vert n\right\rangle$ is the basis for the quantum simple harmonic oscillator Hamiltonian. Furthermore, from references \cite{CohenBook,CohenHarocheDressing1969}, we have
\begin{align}
\overline{\left\langle n_{+}^{\prime }|n_{+}\right\rangle} & =\delta
_{n^{\prime }n}, \\
\overline{\left\langle n_{-}^{\prime }|n_{-}\right\rangle }&=
\delta
_{n^{\prime }n}, \\
\overline{\left\langle n_{+}^{\prime }|n_{-}\right\rangle }&=\left\langle n^{\prime
}\right\vert e^{-\frac{1}{2}\frac{\eta }{\omega }(a-a^{\dagger })}e^{\frac{1%
}{2}\frac{\eta }{\omega }(a^{\dagger }-a)}\left\vert n\right\rangle
=\left\langle n^{\prime }\right\vert e^{\frac{\eta }{\omega }(a^{\dagger
}-a)}\left\vert n\right\rangle \approx J_{n^{\prime }-n}\left( \frac{\gamma B_{1}}{%
\omega }\right),
\end{align}
where $J_{n^{\prime }-n}$ is the Bessel function of the first kind with
order $n^{\prime }-n.$ In references \cite{CohenBook,CohenHarocheDressing1969} they find the energy,
\begin{align}
\frac{E_{n}}{\hbar}=n\omega +\frac{\omega _{0}^{^{\prime }}m}{2}-\frac{\eta ^{2}}{4\omega }, \label{eq:dressedE}
\end{align}
where here, and for the rest of the document, $m=m_z$ with $m=\pm 1$ and $\eta$ is given by
\begin{align}
\eta =\frac{\gamma B_1}{2\lambda ^{1/2}}.
\end{align}
It is then shown in reference \cite{CohenBook,CohenHarocheDressing1969} that the Larmor precession $(\omega _{0}),~$has been effectively scaled $\left( \omega_{0}\rightarrow \omega _{0}^{\prime }\right) $ and in the limit of $\frac{\omega _{0}}{\omega }<<1$ the scaling factor reduces to,
\begin{align}
\omega _{0}^{\prime }=J_{0}\left( \frac{\gamma B_{1}}{\omega }\right) \omega
_{0}, \label{eq:J0}
\end{align}
which we refer to as the $J_0$ approximation. Note that this approximation breaks down when $\frac{\omega _{0}}{\omega }<<1$ is not valid. This is discussed in references \cite{Yabuzaki1974} and \cite{Pinghan2011} and will be addressed below.

The wave function of the system is written as a sum of the basis states weighted by Glauber coefficients, 
\begin{align}
\psi (t)=\sum_{n,m=\pm 1}e^{in\omega t+i\frac{1}{2}m\omega _{0}^{\prime }t}%
\frac{a_{n}}{\sqrt{2}}\overline{\left\vert n,m\right\rangle } \label{eq:wavefun}
\end{align}
Notice that the last term of the energy in equation \ref{eq:dressedE} is a constant energy shift, and thus does
not contribute to the dynamics, this is the reason for its absence in the phase of the wave function in equation \ref{eq:wavefun}. Furthermore, the changes in $\eta $ are
suppressed by a factor $\omega $ and $\lambda ^{\frac{1}{2}}, $ this allows
us to add an energy shift as a perturbation in the Hamiltonian while introducing
negligible error. 
We now present a brief summary of critical dressing and show the results of
optimization within a simulation of critical spin dressing for a system of
neutrons and $^{3}$He.  

\subsection{Critical Dressing}
\label{sec:critDressing}
In the presence of two spin species with different gyromagnetic ratios, e.g. $\gamma_j, \gamma_k$ the
dressing parameters $\left(B_1,\omega \right)$ can be tuned such that the effective Larmor precession ($\omega_0^{\prime}=\gamma^{\prime}B_0$)
is the same for the two species, i.e. $\gamma_j^{\prime} = \gamma_k^{\prime}$. In the $J_0$ approximation this can be achieved if we define $\alpha =\gamma
_{j}/\gamma _{k}$; then, in terms of the dressing parameter $%
x=\gamma B_{1}/\omega $, critical dressing occurs when $x=x_c$ where $x_c$ is found by solving the following equation,

\begin{align}
J_{0}\left( x_{c}\right) =\alpha J_{0}\left( \alpha x_{c}\right). 
\label{eq17}
\end{align}%

However if the $J_0$ approximation is not valid then the effective gyromagnetic ratio can be calculated
numerically to high accuracy by writing the matrix elements of the
Hamiltonian over a large range of $n$ and $n^{\prime }$ and then
diagonalizing the resulting matrix. The resulting eigen values of the
diagonalized matrix will determine the Zeeman splitting where the $n^{%
\mathrm{th}}$ entry is the order of the perturbation. This method is further
discussed in references \cite{Yabuzaki1974} and \cite{Pinghan2011}. Thus the effective frequency can be determined from 
\begin{align}
\omega _{0}^{\prime }=\frac{\Delta E}{\hbar }
\label{eq18}
\end{align}
Where $\Delta E$ is the $n^{\mathrm{th}}$ Zeeman splitting.

First observations of critical spin dressing are discussed in reference \cite{RezaCritObserve2018}, which is a thorough reference for critical spin dressing. Here we present results from Monte-Carlo simulations optimized for
the critical dressing parameters of neutrons and $^{3}$He over a wide
range of $\omega $ and where, in this case, $\alpha = \gamma_3/\gamma_n \approx 1.11$ and $\gamma_3,\gamma_n$ are the $^3$He and neutron gyromagnetic ratios. The optimization was
performed until the effective gyromagnetic ratios differ by less than 1
part in $10^{10}$. It was observed that the $J_0$ approximation is
overall very good for $\omega >> \omega_0$, however, as the dressing frequency approaches the holding
field frequency, differences between the simulation and the $J_{0}~$approximation
can grow large.  Monte-Carlo results are compared to numerically calculated higher order perturbation theory, and the $J_{0}$ approximation in table \ref{table:freqs}. 
\begin{table}[h]
\centering
\begin{tabular}{ccccccc}
\hline
\textbf{$\omega$} (rad s\textsuperscript{-1}) & $\frac{\omega}{\omega_{0}}$ & \textbf{$B_{\text{J}_0}$} ($\mu$T)
& \textbf{$B_{\text{pt}}$} ($\mu$T) & \textbf{$B_{\text{sim}}$} ($\mu$T) & $\frac{B_{%
\text{sim}}-B_{\text{J}_0}}{B_{\text{J}_0}}{\mkern-2mu\times\mkern-2mu}
100\% $ & $\frac{B_{\text{sim}}-B_{\text{pt}}}{B_{\text{pt}}}{\mkern%
-2mu\times\mkern-2mu} 100\%$ \\ \hline
100000 & 164 & 648.8541 & 648.8433379 & 648.8345280 & -0.003 & -1.3578${\mkern%
-2mu\times\mkern-2mu}10^{-3}$ \\ 
30000 & 49.1 & 194.6562 & 194.6202953 & 194.6176552 & -0.020 & -1.3565${\mkern%
-2mu\times\mkern-2mu}10^{-3}$ \\ 
18000 & 29.4 & 116.7937 & 116.7338304 & 116.7322440 & -0.053 & -1.3590${\mkern%
-2mu\times\mkern-2mu}10^{-3}$ \\ 
10000 & 16.4 & 64.8854 & 64.7775066 & 64.7766232 & -0.168 & -1.3637${\mkern%
-2mu\times\mkern-2mu}10^{-3}$ \\ 
6000 & 9.81 & 38.9313 & 38.7511230 & 38.7505920 & -0.466 & -1.3704${\mkern%
-2mu\times\mkern-2mu}10^{-3}$ \\ 
4200 & 6.87 & 27.2519 & 26.9938929 & 26.9935194 & -0.948 & -1.3836${\mkern%
-2mu\times\mkern-2mu}10^{-3}$ \\ 
3000 & 4.91 & 19.4656 & 19.1026874 & 19.1024180 & -1.866 & -1.4100${\mkern%
-2mu\times\mkern-2mu}10^{-3}$ \\ 
2400 & 3.93 & 15.5725 & 15.1162445 & 15.1160267 & -2.931 & -1.4410${\mkern%
-2mu\times\mkern-2mu}10^{-3}$ \\ 
2100 & 3.43 & 13.6259 & 13.1019402 & 13.1017478 & -3.847 & -1.4682${\mkern%
-2mu\times\mkern-2mu}10^{-3}$ \\ 
1800 & 2.94 & 11.6794 & 11.0633252 & 11.0631580 & -5.276 & -1.5115${\mkern%
-2mu\times\mkern-2mu}10^{-3}$ \\ \hline
\end{tabular}%
\caption{Critical dressing parameters for a range of $\omega$ values where $B_{J_0}$ is the solution to critical dressing given by
reference \cite{golub1994}, $B_{\text{pt}}$ the solution found numerically from
higher order perturbation theory, and $B_{\text{sim}}$ is the critical
dressing field found using simulations. The value of $B_{\text{sim}}$ was
optimized such that $\protect\gamma _{n}^{\prime}=\protect\gamma _{\text{%
He}}^{\prime}$ to within 1 part per $10^{10}$. This corresponds to $%
\protect\theta $ changing by less than $5{\mkern-2mu\times \mkern-2mu}10^{-5}
$ rad over 1000 seconds. All the above were optimized using a $B_{0}=3~\mu$T. where, in this case, $\omega'_0 \sim 380$ rad/s.}
\label{table:freqs}
\end{table}

While for estimation purposes the $J_{0}$ approximation is fast and fairly accurate, for precision measurements it may not be accurate enough, for example, in the search for permanent electric dipole moments. 

\subsection{The critically dressed neutron electric dipole moment}

In the absence of spin-dressing the frequency of precession in a uniform magnetic field $(B_{0})~$and
electric field $(E),$ given the existence of a permanent neutron electric dipole moment $(d_{n}),$ with magnetic dipole moment $\mu _{n}~$is given as
\begin{align}
\omega _{0}=\frac{2\mu _{n}B_{0}}{\hbar }+\frac{2d_{n}E}{\hbar }.
\end{align}

However in the case of critical spin dressing the precession frequency will be modified according to Eq. \ref{eq18}. The  experimental signature of a non-zero $d_n$ is a difference in precession frequency for electric and magnetic fields parallel vs. anti-parallel
\begin{align}
\Delta\omega=\omega_{\uparrow\uparrow}^\prime-\omega_{\uparrow\downarrow}^\prime
\end{align}
In the $J_0$ approximation this is given by 
\begin{align}
\Delta\omega=J_0(x_c)\frac{4d_n E}{\hbar}
\end{align}

\begin{figure}
	\begin{center}
		\includegraphics[width=.7\textwidth]{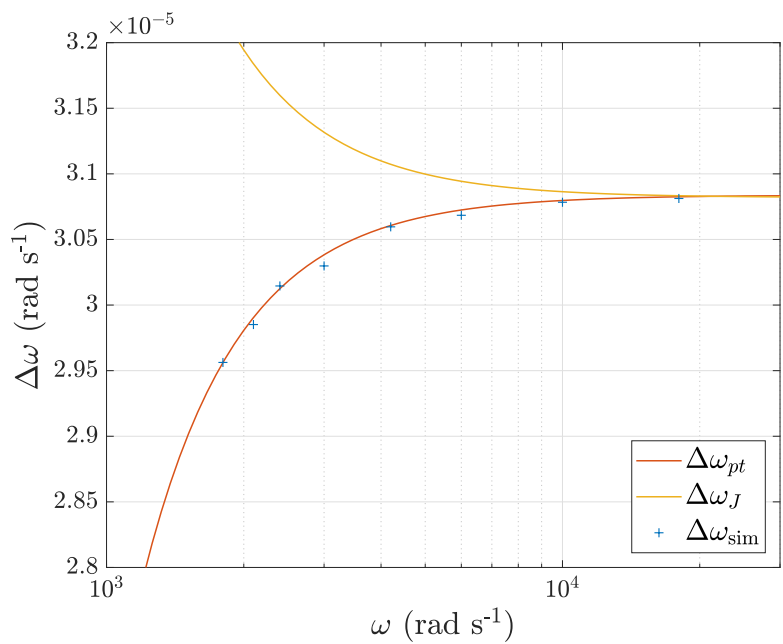}
	\end{center}
	\caption{Frequency shift due to a neutron EDM of $d_n=1\times10^{-25}$~e~cm and 75 kV/cm electric field in a critically dressed system of $^3$He and neutrons. Excellent agreement is observed between the shift predicted from the simulation ($\Delta\omega_{sim}$) and the higher order perturbation theory ($\Delta\omega_{pt}$). The deviation from the simulation and the shift predicted from the $J_0$ approximation ($\Delta\omega_J$) is apparent at lower frequencies. The deviation from a constant for the $J_0$ approximation is due to deviation of $x_c$ from that given by Eq. \ref{eq17}. For these calculations $B_{0}=3~\mu$T giving $\omega'_0 \sim 380$ rad/s}
	\label{fig:hoAgree}
\end{figure}

Beyond the $J_0$ approximation, $\omega_0^\prime$ can be determined numerically as discussed above or from the Monte Carlo simulation. 
We illustrate the onset of the breakdown of the $J_0$ approximation in Fig. \ref{fig:hoAgree} where it is shown that the precession frequency shift determined
from the simulation is best predicted by the numerical diagonalization of a higher order perturbation theory where the shift due to the electric field is
included in the Hamiltonian used to determine the perturbed Zeeman splitting. From
this result we might assume that this numerical method can be used to predict the
relaxation and frequency shifts that are introduced with inhomogeneities in
the holding fields. However this technique will not provide an accurate
prediction, because the frequency shift is an extrinsic effect, it
is determined from the strength of the spectrum for the correlation function
at the Larmor frequency. Since the frequency has been changed by the spin dressing we may also
expect the contribution from the correlation function to change as well. Therefore a more detailed investigation of the relaxation and frequency shifts due
to inhomogeneities of the field is desired; in the following section we
present a model for this purpose.

\section{Relaxation and Frequency shifts from field inhomogeneities in a spin dressed system. }

We assume here that equation \ref{eq:wavefun} is the
solution to the Hamiltonian; for $\omega_0<<\omega$ the $J_0$ approximation is valid. When that condition is broken the exact expansion must be used, or alternately, computed to high accuracy numerically, this is discussed in section \ref{sec:critDressing}. We use this solution to add additional
perturbations in the Hamiltonian, these perturbations take the form of an
electric field $\mathbf{E}=E\hat{z},$ and inhomogeneous field functions for the
spin dressing $B_{1}$ oscillating at a frequency $\omega$, and DC holding field $B_{0}.~$ We will discuss solutions in the particle rest frame such that spatial $B$-field inhomogeneities couple with the particle motion to become time-dependent variations. The perturbation is thus time varying $%
\ \left\langle n\right\rangle =\lambda (t)$.~However, this will be
represented by a change in the field $B_{1}\rightarrow \left\langle
B_{1}\right\rangle +\delta B_{1}(t),$ where for now $\delta B_{1}(t)\ $is
some arbitrary function of time, but is small compared to $\left\langle
B_{1}\right\rangle $. A similar term for the Hamiltonian occurs in reference \cite{OspelkausDressingGradient2008}, where they also model the gradient of a dressing field. Recall that this change is a negligible shift in the
total energy, shifting all levels by the same amount at any instant, thus
the overall dynamics can be well described if we include the term $\frac{%
\gamma \delta B_{1}(t)}{\left\langle \lambda \right\rangle ^{\frac{1}{2}}}%
\frac{\sigma_x}{4}\left( a^{\dagger }+a\right) ,$ this will result in a time dependent energy,
but does not introduce significant error as long as $\gamma \delta B_{1}$ is
small compared to $\omega $ and $\gamma B_{1}.$ \ Interestingly, $\gamma
\delta B_{1}$ need not be small compared to $\gamma B_{0}.$ Furthermore we require $\omega >>\omega _{0}~$ if we wish to use the approximation in equation \ref{eq:J0}. For simplicity we use the notation,
\begin{align}
\omega _{1i}(t)=&\gamma \delta B_{1i}(t),\\
\omega _{0i}(t)=&\gamma \delta B_{0i}(t)+\gamma v_{j}E_{z}/c. \label{eq:wsimp}
\end{align}
were the $i$ index represents the fluctuations due to the field in that direction, and the number index determines the source, a $0$ index indicates it is generated from inhomogeneities of the holding field, while terms with the $1$ index are generated from the dressing field which oscillates at frequency $\omega.$ With this simplification the perturbing Hamiltonian is written as
\begin{align}
H_{pert}& =\frac{1}{4\lambda ^{1/2}}\left[
\omega _{1x}(t)\sigma _{x}+\omega _{1y}(t)\sigma _{y}+\omega _{1z}(t)\sigma
_{z}\right] \left( a+a^{\dagger }\right) \label{eq:Hpert}\\
& +\omega _{0x}(t)\frac{\sigma _{x}}{2}+\omega _{0y}(t)\frac{\sigma _{y}}{2}%
+\omega _{0z}(t)\frac{\sigma _{z}}{2}.\nonumber
\end{align}

Thus, the total Hamiltonian is written as,

\begin{align}
\frac{H}{\hbar }=\omega a^{\dagger }a+\frac{\omega _{1}}{4\lambda ^{1/2}}%
\sigma _{x}\left( a+a^{\dagger }\right) +H_{pert}.\label{eq:H}
\end{align}

We continue to evaluate this Hamiltonian to second order in perturbation
theory for all off-diagonal matrix elements and diagonal elements
oscillating at $\omega ,$ assuming a constant energy for the time evolution. This is equivalent to neglecting the $\omega_{0z}(t) \sigma_z$ term in equation \ref{eq:Hpert}. 
We include the DC field diagonal matrix elements that arise from the $\omega_{0z}(t) \sigma_z$ term in section \ref{sec:T2}, where
time evolution in the energy is required, this is discussed in detail in that section. We will find expressions for
frequency shifts, and transverse relaxation, $T_{2}.$ We do no explicitly consider $T_1,$ because it does not
seem to hold much interest given that we are considering the spin dressed system of a gas, which has typically undergone a $\pi
/2 $ pulse. We will identify the leading terms that contribute to $T_{2}$, and
show that this model is qualitatively in agreement with previous formulations found in
reference \cite{pignol2015,mcgregorrf,CatesHapper1988,mcgregor} considering the effects of spin dressing.

\subsection{Relaxation and frequency shifts in 2$^{\text{nd}}~$%
order perturbation theory.}

We will write the relaxation and frequency shifts in terms of the wave
function determined from perturbation theory. To do this we will find the
expectation value of the spins aligned in the plane perpendicular to the DC
holding field,
\begin{align}
\left\langle \sigma _{x}+i\sigma _{y}\right\rangle =\left\langle \sigma
_{+}\right\rangle.
\end{align}%
From the real part we find $T_{2}$ relaxation and the imaginary part the
accumulated phase due to the frequency shift. Despite deriving the matrix
elements in the representation where the diagonal Pauli Matrix is along $x$,
we write them in terms of a basis diagonal in $z$. We start with the spins in the
plane of Larmor precession, for example after a $\frac{\pi }{2}$ pulse.
Since we are now in the $z$ basis we have, 
\begin{align}
\psi ^{(0)}=\frac{1}{\sqrt{2}}\binom{1}{1}=\frac{1}{\sqrt{2}}\left(
\left\vert +\right\rangle +\left\vert -\right\rangle \right) .
\end{align}%
where to first order the elements of the wave function are, after having
summed over the Glauber states,

\begin{align}
\psi ^{(1)}=-i\int_0^t dt^{\prime }\left\vert f\right\rangle \,\left\langle
f|H|i\right\rangle dt^{\prime },
\end{align}%
the contribution to second order is,

\begin{align}
\psi ^{(2)}=(-i)^{2}\int_0^{t} dt^{\prime }\int_0^{t'} dt^{\prime \prime }\left\vert
f\right\rangle \left\langle f|H|k\right\rangle \,\left\langle
k|H|i\right\rangle dt^{\prime }dt^{\prime \prime },
\end{align}%
so that our total perturbed wave function to second order is,

\begin{align}
\psi =\psi ^{(0)}+\psi ^{(1)}+\psi ^{(2)}.
\end{align}

Our task is the evaluation of the matrix elements of the perturbing
Hamiltonian. However we can be more restrictive of the terms required in our
final calculation if we consider a general perturbed wave function, 
\begin{align}
\psi =\binom{a}{b},
\end{align}%
from this we find the expectation value of $\left\langle \sigma
_{+}\right\rangle $, 
\begin{align}
\left\langle \sigma _{+}\right\rangle =2a^{\ast }b.
\end{align}

Now, $\left\langle \sigma _{+}\right\rangle $ can be written in terms of the
matrix elements used to build $\psi .$ For this notation we write~$%
\left\langle +|H|+\right\rangle =H_{++},$ and $\left\langle
-|H|+\right\rangle =H_{-+}$, etc. The evaluation and discussion of $%
\left\langle \sigma _{+}\right\rangle $ is completed in Appendix \ref{sec:2ndordertheory}, where we find,
\begin{align}
\left\langle \sigma _{+}\right\rangle =1-2t\int_{0}^{t}H_{-+}(0)H_{+-}(\tau
)d\tau -4t\mathrm{Re}\left( \int_{0}^{t}H_{--}(0)H_{--}\left( \tau \right)
d\tau \right).
\end{align}
From this we find the frequency shift, 
\begin{align}
\delta \omega & =\mathrm{Im}\left( \frac{d\left\langle \sigma
_{+}\right\rangle }{dt}\right) =\mathrm{Im}\left(
2\int_{0}^{t}H_{-+}(0)H_{+-}(\tau )d\tau \right), \\
\delta \omega & =2\mathrm{Im}\left( \int_{0}^{\infty }H_{-+}(0)H_{+-}(\tau
)d\tau \right),
\end{align}%
where we assumed the observation time is much longer than the correlation
time in order to change the limit from $t$ to $\infty $. Reference \cite{GolubSteyerlRedfield} assert that the equations are valid for intermediate times, as we
find here, however we will continue with the Fourier transform due to their
simplicity. For the transverse relaxation we have, 
\begin{align}
\frac{1}{T_{2}}=2\mathrm{Re}\left( \int_{0}^{\infty }H_{-+}(0)H_{+-}(\tau
)d\tau \right) + 4\mathrm{Re}\left( \int_{0}^{t}H_{--}(0)H_{--}\left( \tau \right)
d\tau \right).
\end{align}

The matrix elements for all unique terms in the Hamiltonian are evaluated in
the appendix \ref{sec:sigelements}. The results are presented here,

\begin{align}
\left\langle m^{\prime }\right\vert \sigma _{x}\left\vert m\right\rangle &=%
\frac{1}{2}\left( 1-mm^{\prime }\right) e^{i\frac{1}{2}\left( m-m^{\prime
}\right) \omega _{0}^{\prime }t}, \\
\left\langle m^{\prime }\right\vert \sigma _{x}\left( a+a^{\dagger }\right)
\left\vert m\right\rangle &=\lambda ^{\frac{1}{2}}e^{i\frac{1}{2}\left(
m-m^{\prime }\right) \omega _{0}^{\prime }t}\cos (\omega t)(1-mm^{\prime }),
\\
\left\langle m^{\prime }\right\vert \sigma _{y}\left\vert m\right\rangle &=%
\frac{i}{2}e^{i\frac{1}{2}\left( m-m^{\prime }\right) \omega _{0}^{\prime
}t}\left( m-m^{\prime }\right) J_{0}\left( \frac{\gamma B_{1}}{\omega }%
\right)+\left( m+m'\right) \sin (\omega t)J_{1}\left( \frac{\gamma B_{1}}{\omega }\right), \\
\left\langle m^{\prime }\right\vert \sigma _{y}\left( a+a^{\dagger }\right)
\left\vert m\right\rangle &=i\lambda ^{\frac{1}{2}}J_{0}\left( \frac{\gamma
B_{1}}{\omega }\right) e^{i\frac{1}{2}\left( m-m^{\prime }\right) \omega
_{0}^{\prime }t}\cos \left( \omega t\right) \left( m-m^{\prime }\right) ,\\
\left\langle m^{\prime }\right\vert \sigma _{z}\left\vert m\right\rangle &=%
\frac{1}{2}\left( m+m^{\prime }\right) e^{i\frac{1}{2}\left( m-m^{\prime
}\right) \omega _{0}^{\prime }t}-i\left( m-m^{\prime }\right) e^{i\frac{1}{2}\left( m-m^{\prime }\right) \omega _{0}^{\prime }t}J_{1}\left( \frac{\gamma B_{1}}{\omega }\right) \sin (\omega t), \\
\left\langle m^{\prime }\right\vert \sigma _{z}\left( a+a^{\dagger }\right)
\left\vert m\right\rangle &=e^{i\frac{1}{2}\left( m-m^{\prime }\right)
\omega _{0}^{\prime }t}\lambda ^{\frac{1}{2}}J_{0}\left( \frac{\gamma B_{1}}{%
\omega }\right) \cos \left( \omega t\right) \left( m+m^{\prime }\right).
\end{align}

The matrix elements shown above are all in terms of the $J_0$ approximation, if more accuracy is required the dressing factor ($ J_0(\gamma B_1 / \omega)$) should be replaced by a dressing factor found numerically using Eq. \ref{eq18}. For simplicity we write the result as a double sum of possible terms. The
cross terms do not vanish, if there is a field shape that depends on the
same variable, these can be written in terms of the correlation of a
function of the same variable (e.g. $B_{1x}^{\prime }\propto
f(x),~B_{1y}^{\prime }\propto g(x)),$ then the fields will be correlated.
These terms only vanish in special cases,eg. for $B_{1x}^{\prime }\propto x,$
with $B_{1y}^{\prime }\propto x^{2}.$ We ignore cross terms between the DC
and AC fields as these terms always contain some rapidly oscillating phase
and can be neglected, this is shown in the appendix \ref{sec:2ndorderterms}. The relaxation can be
written in terms of modified variables $\omega _{1j}$ $\rightarrow \omega
_{1j}^{\prime },$ specifically, 
\begin{align}
\omega _{1x}^{\prime }\left( t\right) & =\gamma \delta B_{1x}(t), \\
\omega _{1y}^{\prime }\left( t\right) & =i\gamma J_{0}(\left\langle
x\right\rangle )\delta B_{1y}(t), \\
\omega _{1z}^{\prime }\left( t\right) & =\gamma J_{0}(\left\langle
x\right\rangle )\delta B_{1z}(t), \\
\omega _{0x}^{\prime }\left( t\right) & =\gamma \left( \delta B_{0x}(t)+%
\frac{v_{y}(t)}{c^{2}}E\right) , \\
\omega _{0y}^{\prime }\left( t\right) & =i\gamma J_{0}(\left\langle
x\right\rangle )\left( \delta B_{0y}(t)+\frac{v_{x}(t)}{c^{2}}E\right) , \\
\omega _{0z}^{\prime }\left( t\right) & =\gamma J_{0}(\left\langle
x\right\rangle )\delta B_{0z}(t), \\
\omega _{0zq}^{\prime }\left( t\right) & =\gamma J_{1}(\left\langle
x\right\rangle )\delta B_{0z}(t)).
\end{align}%
Where, $\omega _{0zq}^{\prime }$ corresponds to the first harmonic of the oscillating terms in the matrix elements, which arise due to the dressing field, this is discussed in the appendix \ref{sec:2ndorderterms}. Notice that $\omega _{1x}^{\prime }(t)=\omega _{1x}(t),~$because $x$
corresponds to the direction of the applied dressing field, it does not obtain a factor $J_0(x)$. With these
definitions we can write our phase shift as,

\begin{align}
\delta \omega & =2\mathrm{Im}\left( \frac{1}{8}\sum_{k=x,y}\sum_{j=x,y}%
\int_{0}^{\infty }\omega _{1j}^{\prime }(0)\omega _{1k}^{\prime }(\tau )\cos
(\omega \tau )e^{-i\omega _{0}^{\prime }\tau }d\tau \right)  \\
& +2\mathrm{Im}\left( \frac{1}{4}\sum_{k=x,y}\sum_{j=x,y}\int_{0}^{\infty
}\omega _{0j}^{\prime }(0)\omega _{0k}^{\prime }(\tau)e^{-i\omega _{0}^{\prime
}\tau }d\tau \right) 
\end{align}%
Notice that there is no cross correlation between the AC and DC terms, $\omega_{1i}$ and $\omega_{0k}$, in the frequency shift. This is because a quickly oscillating phase appears in all cross correlation terms in the expansion, this is shown in Appendix \ref{sec:2ndorderterms}. Thus, there is no linear in $E$ frequency shift generated by gradients in the dressing field. Typically the first term(s) can be ignored, and the second term(s) dominates
the frequency shift. We turn our attention to the transverse relaxation
where we hold off on writing the zero frequency term for the DC field
inhomogeneity. This is investigated in section \ref{sec:T2}, where we show
that for the zero frequency part of the DC relaxation we must include the
time dependent energy in the Hamiltonian. The current method does not take
into account the time variation of the energy. This fails for the diagonal DC field component ($\omega _{0z}^{\prime }$), because $\omega
_{0z}^{\prime }(t)$ is a relatively large contribution to the overall phase.
In fact, given that this is a static contribution to the diagonal matrix
element, it is the whole contribution of the phase for that term, and while
it is small compared to $\left\langle \omega _{0}\right\rangle $, $%
\left\langle \omega _{1}\right\rangle $ and $\omega $, it cannot be ignored when it is the sole contribution to the phase. For the off-diagonal terms, or diagonal terms
oscillating at $\omega $, the energy dependence of the time evolution
operator can be ignored due to the negligible size of the phase shift compared
to $\left\langle \omega _{0}^{\prime }\right\rangle $, and $\omega $.
Starting from the evaluation of $\left\langle \sigma _{+}\right\rangle $ in
terms of the wave function constructed from 2$^{\text{nd}}$ order
perturbation theory we find, (after omitting $\omega _{0z}^{\prime },$)

\begin{align}
\frac{1}{T_{2}}& =2\mathrm{Re}\left( \frac{1}{4}\sum_{k=x,y}\sum_{j=x,y}%
\int_{0}^{\infty }\omega _{1j}^{\prime }(0)\omega _{1k}^{\prime }(\tau )\cos
(\omega \tau )e^{-i\omega _{0}^{\prime }\tau }d\tau \right)  \\
& +2\mathrm{Re}\left( \frac{1}{4}\sum_{k=x,y}\sum_{j=x,y}\int_{0}^{\infty
}\omega _{0j}^{\prime }(0)\omega _{0k}^{\prime }(\tau )e^{-i\omega
_{0}^{\prime }\tau }d\tau \right)  \\
& +4\mathrm{Re}\left( \frac{1}{2}\int_{0}^{\infty }\omega _{1z}^{\prime
}(0)\omega _{1z}^{\prime }(\tau )\cos \left( \omega \tau \right) d\tau
\right)\\
&+\mathrm{Re} \left(i\int_0^\infty e^{-i\omega _{0}^{\prime }\tau }\sin (\omega \tau )\omega' _{1x}(0)\omega'_{0zq}(\tau )d\tau\right) \label{eq:qpm1Cross} \\
&+\mathrm{Re}\left(i\int e^{-i\omega_{0}^{\prime }\tau }\sin (\omega \tau )\omega' _{0zq}(0)\omega'_{0zq}(\tau )d\tau\right). \label{eq:qpm1Square}
\end{align}%
In general, there is no reason a field in one direction will not be
correlated with another over the same position coordinate. For example,
Maxwell's equation does not preclude the relation $\frac{dB_{x}}{dy}\propto 
\frac{dB_{y}}{dy}$, in this case these two field inhomogeneities would be
correlated due to a similar dependence on the $y$ position variable, and
thus the cross terms must be included for an accurate prediction of the
relaxation. Furthermore, we will not consider the terms found in equation \ref{eq:qpm1Cross} and \ref{eq:qpm1Square} any further. These terms contribute as the difference in the real part of the spectrum at $\omega\pm\omega'_0$, and are highly suppressed for typical spin dressing parameters, when $\omega>>\omega'_0$, but we included them here to maintain generality.

The transverse relaxation that we find is nearly in agreement with the
formulation found in reference \cite{mcgregorrf}, with the exception that our contribution from the $y$ and $z$ field power spectrum contain an extra factor $J_{0}^{2}\left( \left\langle x\right\rangle \right) $ due to dressing. We now finish the derivation of $T_{2}$ by examining terms in the diagonal component of the Hamiltonian containing $\omega _{0z}$ without assuming a dressed energy that is constant in time. We will find that diagonal terms typically dominate the relaxation, this is because typically the magnitude of the gradient for the on-axis field component is similar to the off-axis component, so that all other terms can be ignored. However, we should be aware that there are solutions to Maxwell's equation where this assumption is not valid. 

\subsection{T$_2$ due to spatial inhomogeneities in a spin dressing and holding
field.}\label{sec:T2}
Now we create a model to determine transverse relaxation ($T_{2}$) that
incorporates diagonal heterogeneities in the spin dressing field. This is not considered in the previous section or in reference~\cite{mcgregorrf}, where the AC gradients that cause relaxation spatially average to zero. The model requires that we  incorporate our time dependence into the energy of the state determined from the Hamiltonian, given by 
\begin{align}
H=\omega _{z}(t)\frac{\sigma _{z}}{2}.
\end{align}
 Here $\omega _{z}=\gamma J_{0}(x(t))\left( \left\langle
B_{0}\right\rangle +\delta B_{0}(t)\right) $ and the time evolution operator
is given as,
\begin{align}
U(\delta t,0)=\exp \left\{ -i\gamma J_{0}\left( x\left( t\right) \right) 
\left[ \left\langle B_{0}\right\rangle +\delta B_{0}(t)\right] \frac{\sigma
_{z}}{2}\delta t\right\}.
\end{align}
Alternatively one could proceed with the propagator given In reference \cite{BevilacquaDressing}, where they  in reference \cite{Yuen2018} to simultaneously account for a distribution of frequencies, or in reference  we do not consider this because, typically for systems investigated here the precision and stability of the frequency is far greater than the field uniformity. The general solution is,
\begin{align}
\left\vert \alpha \right\rangle =c_{+}\left\vert +\right\rangle
_{z}+c_{-}\left\vert -\right\rangle _{z}.
\end{align}
We start with the spins along the $+x$ direction, equivalent to a system
directly after a $\frac{\pi }{2}$ pulse, our solution becomes,
\begin{align}
\left\vert \alpha \right\rangle =\frac{1}{\sqrt{2}}\left[ \exp \left(
-i\gamma \int_{0}^{t}J_{0}(x(t^{\prime }))B_{0}(t^{\prime })dt^{\prime
}\right) \left\vert +\right\rangle +\exp \left( i\gamma
\int_{0}^{t}J_{0}(x(t^{\prime }))\delta B_{0}(t^{\prime })dt^{\prime
}\right) \left\vert -\right\rangle \right] .
\end{align}%
To find an expression for $T_2$ we evaluate $\left\langle \sigma _{x}+i\sigma
_{z}\right\rangle =\left\langle \sigma _{+}\right\rangle ,$ and find its
rate of decay,
\begin{align}
\left\langle \alpha ,t\right\vert \sigma _{+}\left\vert \alpha
,t=0\right\rangle & =\frac{1}{2}\left( \left\langle -\right\vert \exp \left(
-i\gamma \int_{0}^{t}J_{0}(x(t^{\prime \prime }))\left( B_{0}(t^{\prime
\prime })\right) dt^{\prime \prime }\right) +\left\langle +\right\vert \exp
\left( i\gamma \int_{0}^{t}J_{0}(x(t^{\prime \prime }))\left( \delta
B_{0}(t^{\prime \prime })\right) dt^{\prime \prime }\right) \right) \nonumber \\
& \times \left\vert +\right\rangle \left\langle -\right\vert \left( \exp
(-i\gamma \int_{0}^{t}J_{0}(x(t^{\prime }))\left( B_{0}(t^{\prime })\right)
dt^{\prime })\left\vert +\right\rangle +\exp \left( i\gamma
\int_{0}^{t}J_{0}(x(t^{\prime }))\left( \delta B_{0}(t^{\prime })\right)
dt^{\prime }\right) \right) \left\vert -\right\rangle. \label{eq:dcT2sig}
\end{align}
We concentrate on the phase in the exponential. To simplify the time dependence in the Bessel function we write,
\begin{align}
\int d\omega _{\phi }& =\gamma \int d\left\{ J_{0}\left( \frac{\gamma
\left\langle B_{1}\right\rangle +\omega _{1x}}{\omega }\right) \left[
B_{0}+\delta B_{0z}(t)\right] \right\},  \\
\omega _{\phi }(t)& =\gamma \int J_{1}\left( \frac{\gamma \left\langle
B_{1}\right\rangle }{\omega }\right) \frac{d\omega _{1x}}{\omega }\left(
B_{0}+\delta B_{0z}(t)\right) +\gamma \int J_{0}\left( \frac{\omega _{1}}{%
\omega }\right) d\delta B_{0z}(t), \\
\omega _{\phi }(t)& \simeq \frac{\gamma }{\omega }\int J_{1}\left( \frac{\gamma
\left\langle B_{1}\right\rangle }{\omega }\right) \left( B_{0}d\omega
_{1x}+\delta B_{0z}(t)d\omega _{1}^{\prime }\right) +\gamma \int J_{0}\left( 
\frac{\gamma \left\langle B_{1}\right\rangle }{\omega }\right) d\delta
B_{0z}(t), \\
\omega _{\phi }(t)& \simeq \frac{\gamma }{\omega }\int J_{1}\left( \frac{%
\gamma \left\langle B_{1}\right\rangle }{\omega }\right) B_{0}d\omega
_{1x}+\gamma \int J_{0}\left( \frac{\gamma \left\langle B_{1}\right\rangle }{%
\omega }\right) d\delta B_{0z}(t), \\
\omega _{\phi }(t)& \simeq \frac{\gamma }{\omega }J_{1}\left( \frac{\gamma
\left\langle B_{1}\right\rangle }{\omega }\right) B_{0}\omega _{1x}+\gamma
J_{0}\left( \frac{\gamma \left\langle B_{1}\right\rangle }{\omega }\right)
\delta B_{0z}(t).
\end{align}
We substitute this back into equation \ref{eq:dcT2sig},
\begin{align}
\left\langle \sigma _{+}\right\rangle =\exp \left[ i\int_{0}^{t}\left( \frac{\gamma }{\omega }J_{1}\left( \frac{\gamma \left\langle
B_{1}\right\rangle }{\omega }\right) B_{0}\omega _{1x}\left( t^{\prime
}\right) +\gamma J_{0}\left( \frac{\gamma \left\langle B_{1}\right\rangle }{%
\omega }\right) \delta B_{0}(t^{\prime })\right) dt^{\prime }\right]. \label{eq:genT20}
\end{align}
This is the general form of the zero frequency decay component of $T_{2}$.
Theoretically if we know $B_{0}(t)~$then the solution is known. However,
perfect knowledge of $B_{0}(t)$ for every particle is challenging.  Thus, we
describe thermal motion by a conditional probability distribution function, a model
that is appropriate for this derivation is described by references \cite{swank,Swank2016}. The evaluation of $\left\langle\sigma_+\right\rangle$ is completed in the appendix \ref{sec:T2dc}, and diagonal DC field contribution to $T_2$ is
\begin{align}
\frac{1}{T_{2~\mathrm{DC}}}& =\mathrm{Re}\left\{S_{\omega'
_{0z}\omega' _{0z}}(0) \right.\\
& +J_{1}(x)^{2}\frac{\omega _{0}^{2}}{\omega ^{2}}S_{\omega _{1x}\omega
_{1x}}(0) \\
& -2J_{1}\left( x\right) \frac{\omega _{0}}{\omega }S_{\omega _{1x}\omega
_{0z}^{\prime }}(0)\left.\right\}. \label{eq:dcT2} 
\end{align}%

Again, we use $\omega _{0}^{\prime }=\gamma J_{0}\left( x\right) B_{0},$
where $S_{\omega _{0}^{\prime }\omega _{0}^{\prime }}(0)$ is the zero
frequency of the spectrum of the autocorrelation function of $\omega
_{0}^{\prime }\left( t\right) .$ In general, we define the spectrum to be,
\begin{align}
S_{\omega _{ij}\omega _{kl}}\left( \omega \right) =\int_{0}^{\infty }\left\langle\omega
_{ij}\left( 0\right) \omega _{kl}(\tau ) \right \rangle e^{-i\omega \tau }d\tau.
\end{align}%
Defining the spectrum from $t=0\rightarrow \infty $,~allows us to keep the imaginary parts, which are necessary for the prediction of the frequency shift. 

The DC diagonal rate is summed with the off-diagonal and AC diagonal rates for the full prediction of $T_{2}$,
\begin{align}
\frac{1}{T_{2}} =\mathrm{Re}& \left\{   \frac{1}{4}\sum_{k=x,y}\sum_{j=x,y}\left[
S_{\omega _{1j}^{\prime }\omega _{1k}^{\prime }}(\omega +\omega _{0}^{\prime
})+S_{\omega _{1j}^{\prime }\omega _{1k}^{\prime }}(\omega -\omega
_{0}^{\prime })\right] \right. \label{eq:rfterms}  \\
& +\frac{1}{2}\sum_{k=x,y}\sum_{j=x,y}S_{\omega _{0j}^{\prime }\omega
_{0k}^{\prime }}(\omega _{0}^{\prime }) \nonumber \\
& + S_{\omega _{1z}^{\prime }\omega _{1z}^{\prime }}(\omega ) \nonumber \\
& +S_{\omega _{0}^{\prime }\omega _{0}^{\prime }}(0) \nonumber \\
& +J_{1}(x)^{2}\frac{\omega _{0}^{2}}{\omega ^{2}}S_{\omega _{1x}\omega
_{1x}}(0) \nonumber \\
&  \left.-2J_{1}\left( x\right) \frac{\omega _{0}}{\omega }S_{\omega _{1x}\omega
_{0z}^{\prime }}(0) \right\}. \nonumber
\end{align}

In the case of the cross correlation $S_{\omega _{1x}\omega _{0z}^{\prime
 }}(0)$ we will find that if the variation in $\omega _{1x}$ and $\omega
_{0z}^{\prime }$ is due to spatial inhomogeneities then only terms that are
a function of the same variable will be correlated. For example if there is
a spin dressing field gradient along $x$, and the static field gradient
along $x$ then the cross correlation will not vanish, $\left( S_{\omega
_{1}^{\prime }\omega _{0}^{\prime }}(0)\neq 0\right)$. This formulation of
the relaxation offers the curious ability to make the cross term the
opposite sign as the squared terms, implying that one can slow down and
essentially cancel the DC relaxation as long as the gradients in each field are
of the right proportion. The spin dressing gradient  that allows this cancellation, i.e. $\frac{1}{%
T_{2}}\approx 0,$ corresponds to, 
\begin{align}
G_{1xj}=G_{0zj}\frac{\omega }{\omega _{0}}\frac{J_{0}(x)}{J_{1}(x)},\text{ for }%
\frac{1}{T_{2}}\approx 0. \label{eq:norelax}
\end{align}%
where, $G_{1xj}=\frac{dB_1x}{dj}$ and $j$ can be $x,y,$ or $z$, and for the DC field gradient, $G_{0zj}=\frac{dB_0z}{dj}.$ Physically, the gradient in the spin dressing field changes the effective gyromagnetic ratio in the correct proportion to the gradient in the holding field allowing the spins to have the same effective frequency throughout the whole volume. When the spins have the same effective frequency across the spatial volume the relaxation from these terms vanishes. Matching the effective gyromagnetic precession across the volume to extend coherence times has recently been reported in reference \cite{BevilaqueGradientCancel}, where they independently predict and experimentally verify that a spin dressed inhomogeneity can counteract an applied DC inhomogeneity to recover the original coherence times, even in the presence of large gradients. Therefore this result is not limited to the Redfield regime as presented here, and is valid where the signal decay becomes dominated by the reversible process of gradient dephasing. We expand on this and claim that from Eq. \ref{eq:rfterms} it is clear that the attenuation of the polarization due to the motion of the spins is also suppressed, making the technique technically superior when compared to refocusing the spins with spin echo. However, some relaxation is unavoidable, because the AC relaxation terms, specifically from the first line in equation \ref{eq:rfterms}, are dominant when the cancellation is sufficient.

\subsection{Comparison to simulations}

Monte Carlo simulations coupled with a 5th order Runge-Kutta integrator are compared to the theoretical predictions for the relaxation and frequency shifts of the spin dressed system. The simulation package is further described in references \cite{riccardoThesis,SchmidERelax2008}. The simulation is for a system of $^{3}$He in a superfluid $^{4}$He bath below 450 mK. The volume is confined to a
rectangular cell of dimensions, $(x,y,z),$ $40\times 10.2\times 7.6$ cm.

\begin{figure}
\begin{center}
\includegraphics[width=.7\textwidth]{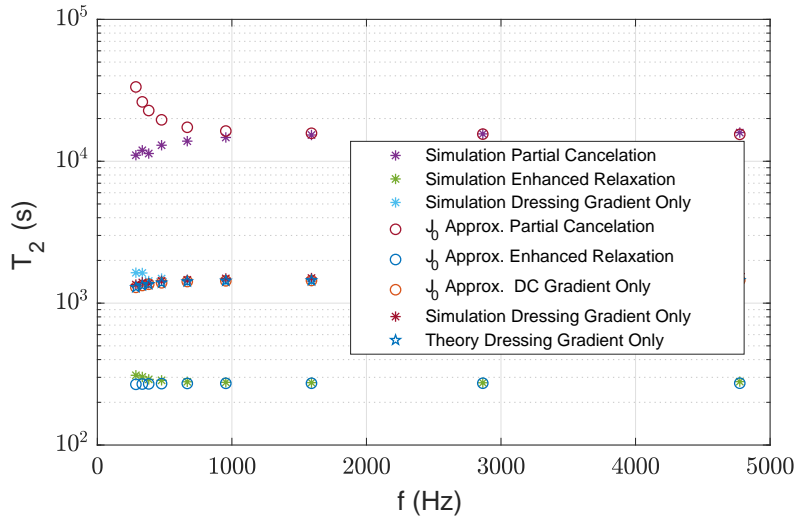}
\end{center}
\caption{$T_2$ in the $J_0$ approximation due to a spin dressing gradient, showing partial cancellation,
3\% effective remnant gradient according to equation \ref{eq:norelax}. The theory points are circles and the simulation data are stars,  results agree until the Bessel function approximation breaks down due to relatively slow AC frequency
compared to the Larmor frequency. The gradient in the uniform field is $1\times10^{-5}~B_0/cm$ in the $z$ direction and $B_0 = 3~\mu$T}
\label{fig:T2SD}
\end{figure}

\begin{figure}
\begin{center}
\includegraphics[width=.7\textwidth]{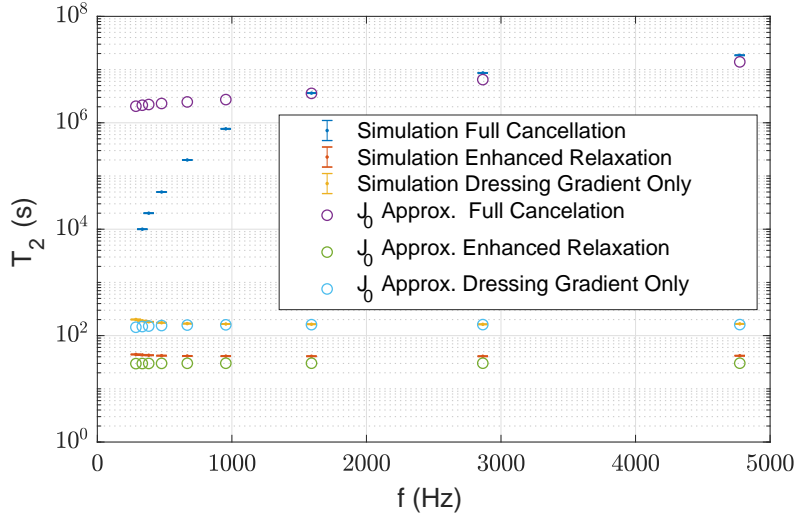}
\end{center}
\caption{$T_2$ in the $J_0$ approximation due to a spin dressing gradient, showing full theoretical
cancellation. Expectedly, full theoretical cancellation does not result in an infinite $T_2$ as the AC terms in the relaxation become the dominant source of relaxation. Otherwise, results agree within the spin dressing $J_0(\omega_1/\omega)$ approximation, the deviation observed is due to this approximation breaking down. The gradient in the uniform field is $3\times10^{-5}~B_0/cm$ in the $z$ direction, and the AC gradient is determined from equation \ref{eq:norelax}.}
\label{fig:nodecay}
\end{figure}

The vanishing relaxation effect, simulated with gradients in both the holding and dressing fields
simultaneously, is shown in figure \ref{fig:T2SD}. In this case the spin dressing gradient is tuned to partially cancel the holding field gradient, with 3\% residual gradient remaining. Full cancellation of the zero frequency terms contributing to the relaxation, is shown in in figure \ref{fig:nodecay}. As expected, total cancellation is not observed as the AC field terms become dominant. The cross term that extends the relaxation time can also enhance the relaxation rate so that relaxation is faster than the sum of the individual rates. The enhanced relaxation is shown in figure \ref{fig:T2SD}, and \ref{fig:nodecay}, where the gradient in equation \ref{eq:norelax} has the opposite sign. Deviation from the simulations is observed in both figure \ref{fig:T2SD}, and \ref{fig:nodecay} as the dressing frequency decreases. This is due to the approximation in equation \ref{eq:J0} becoming invalid. Presumably this discrepancy is largely removed by using the full expansion found in reference \cite{CohenHarocheDressing1969}, or by finding it numerically as described in section \ref{sec:critDressing}. The slope of the $J_0$ approximation around the critical dressing parameter $x_c$ is also required. Nonetheless, a substantial gain in $T_{2}$ is observed.

This effect allows the ability to change the relaxation rate by either tuning the dressing field gradient or the holding field gradient. Typically
for dressed systems we have $\frac{\omega }{\omega _{0}}>\frac{J_{0}(x)}{%
J_{1}(x)}.$ Thus, a change in the dressing field gradient will change the relaxation at a slower rate compared to the holding field gradient.  Therefore the resolution for changing the relaxation is smaller for the dressing field gradient, making it technically easier to manipulate the relaxation through manipulation of that gradient. However, it requires tuning an AC gradient to be in phase with the original dressing field. With modern timing resolution this problem is certainly solvable, but achieving the stability required may be challenging. Furthermore, we can determine the size of the gradient in the spin dressing or holding field by keeping the gradient in one constant and varying  the gradient of the other. 

\begin{figure}[tbp]
	\begin{center}
		\includegraphics[width=.7\textwidth]{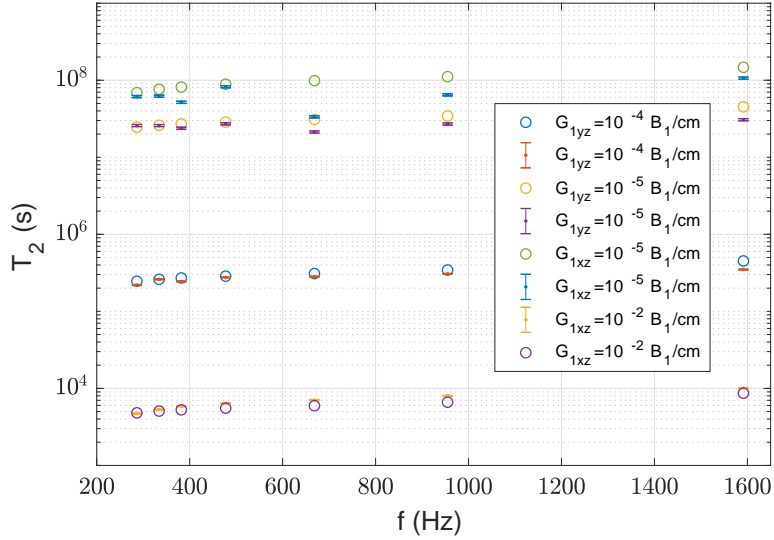}
	\end{center}
	\caption{Comparison of $T_2$ due to gradients in the off-axis components of the spin dressing field for the numerically solved dressing factor versus the simulation. Each gradient was simulated across a range of dressing frequencies. The circles indicate theory prediction and the error bars are from the simulation. There is no deviation observed at the lower frequencies because the dressing factor was computed numerically for the simulation and theory, rather than using the $J_0$ approximation.}
	\label{fig:SDT2rf}
\end{figure}

Typically, the zero frequency components of the field will dominate the
relaxation, however there are solutions to Maxwell's equation where this is
not the case. Thus, simulations for gradients in $\omega _{1x}$ and $\omega
_{1y}$, were performed and compared to the theory. This is shown in figure \ref{fig:SDT2rf}.

For the simulations of the phase shift we include an electric field $\mathbf{%
E}=E~\hat{z}$, and include a magnetic gradient in the $z$ direction with 
the negative of the $x$ direction, $G_{0zz}=-G_{0xx}$. This example scenario is
simple but realistic since it satisfies Maxwell's equations, a condition
that can broken in a simulation. This gradient is efficient in examining
the model as it will give us the minimum number of similarly behaving terms,
and thus the most direct in comparison to simulations. For this case we have the following terms in the Hamiltonian that must be considered,
\begin{align}
\omega _{1z}^{\prime }(t)& =\gamma J_{0}(\left\langle x\right\rangle
)G_{1zz}z(t), \\
\omega _{1x}^{\prime }(t)& =\gamma G_{1xx}x(t), \\
\omega _{0z}^{\prime }& =\gamma J_{0}(\left\langle x\right\rangle )G_{0zz}z(t), \\
\omega'_{0x}& =\gamma \left( G_{0xx}x(t)+\frac{v_{y}(t)}{c^{2}}E\right),  \\
\omega _{0y}^{\prime }\left( t\right) & =i\gamma J_{0}(\left\langle
x\right\rangle )\frac{v_{x}(t)}{c^{2}}E.
\end{align}

For this example we ignore contributions of $\frac{E^{2}}{c^{4}}$ as
typically these are negligible, although we should be cautious in
simulations not to make $E$ large enough to become the dominant term when
searching for other effects. Typically by making $E$ un-physically large we
magnify the desired linear in $E$ effect (for simulations purposes), but it
can be obscured if the relaxation and frequency shifts due to $E^{2}$ become
dominant. If we assume that $E^{2}$ terms can be neglected, and keeping only
terms linear in $E$ we find,

\begin{align}
\delta \omega & =\frac{\gamma ^{2}}{c^{2}}EG_{0xx}J_{0}(\left\langle
x\right\rangle )\mathrm{Re}\left( \int_{0}^{\infty }x(0)v_{x}(\tau
)e^{-i\omega _{0}^{\prime }\tau }\right),  \\
\delta \omega & =-\frac{\gamma ^{2}}{c^{2}}EG_{0xx}J_{0}(\left\langle
x\right\rangle )\left[ \omega _{0}^{\prime }\mathrm{Im}\left(
\int_{0}^{\infty }x(0)x(\tau )e^{-i\omega _{0}^{\prime }\tau }\right) +\frac{%
L_{x}^{2}}{12}\right] , \\
\delta \omega & =-\frac{\gamma ^{2}}{c^{2}}EG_{0xx}J_{0}(\left\langle
x\right\rangle )\left\{ \omega _{0}^{\prime }\mathrm{Im}\left[ \left(
S_{xx}(\omega \right) \right] +\frac{L_{x}^{2}}{12}\right\} .
\end{align}

This was simulated for the range of critically dressed frequencies found in table \ref{table:freqs}. We find good agreement that is consistent with the combined error of the simulation and extraction of the frequency shift. 

\begin{figure}[tbp]
\begin{center}
\includegraphics[width=.7\textwidth]{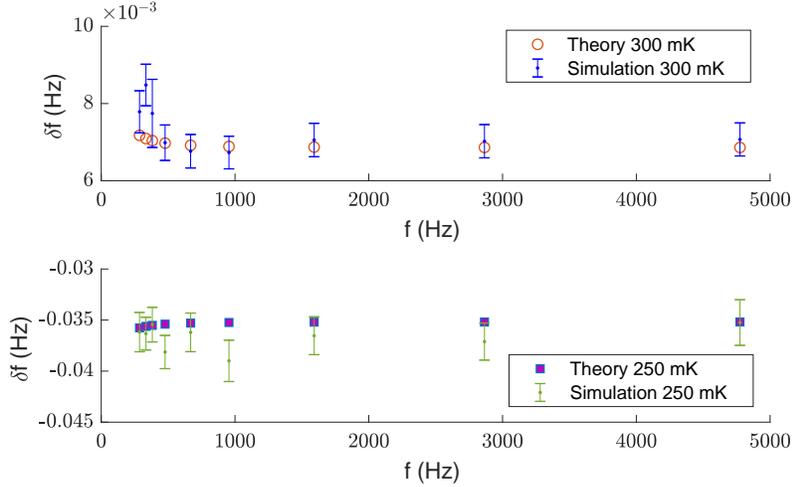}
\end{center}
\caption{Linear in E frequency shift in a rectangular cell of $%
0.4~\times~0.102~\times~0.076$~m (x,y,z) dimensions respectively. The spins
contained in the cell are under the influence of a dressed field. The upper
plot is modeling $^3$He in a superfluid helium-II bath at 250~mK in a 2~$%
\protect\mu$T holding field along the $z$ direction. The lower plot is for
the same system 300~mK in a 3~$\protect\mu$T holding field. The E field is
750~kV/cm along the $z$ direction and the DC field gradient is $%
2~\mu$T/m along the $x$ direction, this gradient is large compared to gradients achieved in laboratory fields in order to increase the effect above the resolution of the simulation. In this simulation the EDM is set to zero. }
\label{fig:SDdf}
\end{figure}

\bigskip We now examine the effect of modulation of the spin dressing
parameters on the phase shifts and relaxation. 

\subsection{Modulated critical dressing}

Modulated critical spin dressing is the modulation of parameters of the critically dressed spin system around the critically dressed value. It is a technique that can be utilized to
optimize statistical precision and mitigate systematic drifts. Experimental observation of the systematic improvement is presented in \cite{RezaCritObserve2018}. In reference \cite{bradfSensitivity} they find pulsed modulation to be the optimum modulation technique for statistical sensitivity.  In pulse modulated dressing the gyromagnetic
frequencies of the two spin species are momentarily allowed to shift to a
large difference so that a known angle $\theta _{m}$ is accumulated between
the spins. When $\theta _{m}$ is accumulated an observation period of
critical dressing is continued for a given time. After this time has been
completed the frequency change is reversed and the angle $\theta _{m}$ is
undone. It is shown in reference \cite{bradfSensitivity,DTFnEDM} that a
statistically optimum value for the angle $\theta _{c}$ between the spins
exists. The modulated dressing is used to swing neutron and $^{3}$He spins
to $\pm \theta _{c}$. For example, this is achieved at the start of the
critical dressing when the initial phase between the spin species is equal
to $-\theta _{c}$, after the critical dressing observation period the
modulation pulse is applied where $\theta _{m}=2\theta _{c}$ bringing the
phase to $\theta _{c}$, after another observation period a pulse $\theta _{m}=-2\theta _{c}$ is applied. The pulse train repeats for the duration of the
measurement. For more details refer to reference \cite{golub1994,DTFnEDM}.

\begin{figure}[h!]
\begin{center}
\includegraphics[width=.7\textwidth]{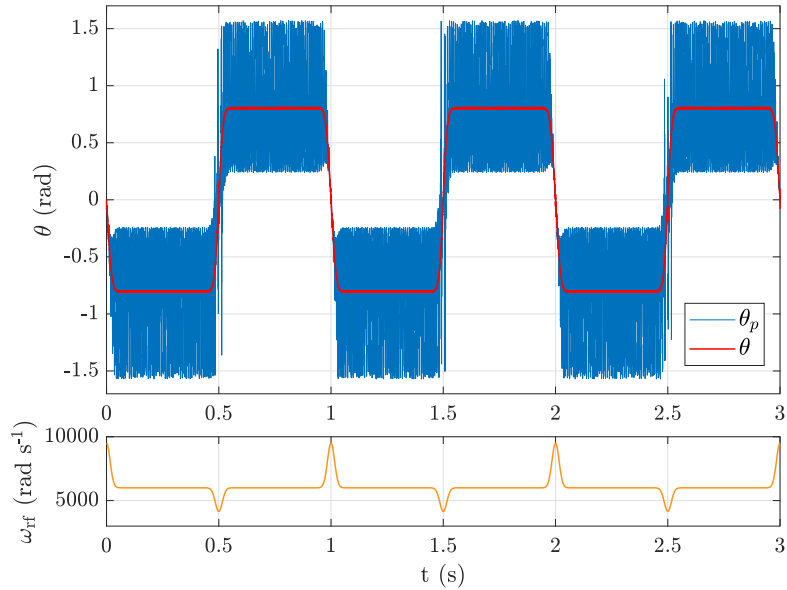}
\end{center}
\caption{Plot of $\protect\theta$, $\protect\theta_p$ and $\protect\omega_{%
\text{var}}$ against time for three complete modulation cycles where
parameters are given by those in Table~\protect\ref{table:dressingpulse} for 
$\protect\omega_{c} = 10000$ rad s\textsuperscript{-1} and $\protect\phi_{%
\text{mod}} = \protect\pi/2$. $\protect\theta$ is the total angle between
the two species' spins and $\protect\theta_p$ is the angle between the
projection of the two species' spins on the plane perpendicular to $B_0$.}
\label{fig:6000_3cycle}
\end{figure}

The modulation technique was optimized across a number of different
modulation strategies and pulses. It is clear from modern timing resolution
that frequency modulation is most precise experimentally. Initially square
pulses in the frequency $\omega $ of the pulse were pursued as the
modulation technique, however this was repeatedly shown to have poor
coherence times. Despite great effort, a set of parameters that approached
acceptable behavior was never determined for the square pulses, despite an analytic solution in reference \cite{golub1994}. It was found
that if the square pulses where smoothed the modulation of frequency would
remain coherent, and ultimately no relaxation or distortion was observable
after $10^{3}$ modulation cycles. Of the functions studied the parametric
function that demonstrates the best performance is,
\begin{align}
\omega_{\mathrm{var}}(r)=\omega+\omega_{\mathrm{amp}}f(r) \label{eq:omegarf}
\end{align}%
where,
\begin{align}
f(r)=\frac{1}{a}e^{-n(r(t)-\frac{\pi }{2})^{2}}-ae^{-n(r(t)-\frac{3\pi }{2}%
)^{2}}, \label{eq:fpulse}
\end{align}%
and, 
\begin{align}
r(t)=\mathrm{mod}_{2\pi }\left( \omega _{\text{fm}}t+\phi _{m}\right), \label{eq:rdef}
\end{align}%
where $\omega _{\text{amp}}$ is the amplitude and $\omega _{\text{fm}}$ is the frequency of the modulation (about 1 Hz), 
$\phi _{m}$ is a modulation phase, the parameter $a$ controls the relative
heights of the positive and negative modulation pulses, and $n$ controls the
sharpness of the peaks. The integral of this pulse can be written in closed
form with the use of the error function, which is a well tabulated function. Using this modulation pulse the required parameters of
the pulse for the desired effect on the spin solution can be calculated
accurately, and quickly, leaving the possibility of feedback timing
corrections during the measurement. A table for the modulation parameters
for a range of $\omega $ is shown in table \ref{table:dressingpulse}. A time
sequence plot of the pulse train for a particular frequency ($\omega =10,000$%
) and the response of the spins are shown in figure \ref{fig:6000_3cycle}.

\begin{table}[h!]
\centering
\begin{tabular}{lllll}
\hline
\textbf{$\omega$} (rad s\textsuperscript{-1}) & \textbf{$\omega_{\text{amp}}$%
} (rad s\textsuperscript{-1}) & \textbf{$a$} & $\theta_0$ (rad) & $%
\sqrt{\langle\Delta\theta^2\rangle}$ (rad) \\ \hline
30000 & 12711.5053 & 0.7256 & 0.7998 & 0.012 \\ 
18000 & 7624.44225 & 0.7257 & 0.7998 & 0.012 \\ 
10000 & 4231.97665 & 0.7260 & 0.7998 & 0.012 \\ 
6000 & 2542.33540 & 0.7260 & 0.8022 & 0.013 \\ 
4200 & 1766.76950 & 0.7280 & 0.7999 & 0.014 \\ 
3000 & 1255.05106 & 0.7300 & 0.8012 & 0.015 \\ 
2400 & 994.364100 & 0.7330 & 0.8006 & 0.018 \\ 
2100 & 862.250665 & 0.7357 & 0.7996 & 0.019 \\ 
1800 & 730.012950 & 0.7395 & 0.7995 & 0.020 \\ \hline
&  &  &  & 
\end{tabular}%
\caption{Optimized parameters for range of $\protect\omega_{c}$ values for
the modulation function given in (\ref{eq:fpulse}) and (\ref{eq:rdef}). For all frequencies $n = 51$ was used and $B_{\text{rf}}$ values
are the same as the $B_{\text{sim}}$ values found in Table~\protect\ref%
{table:freqs}. $\protect\theta_0$ is the time average value of $\protect%
\theta$ between modulation pulses and $\sqrt{\langle\Delta\theta^2\rangle}$ is the rms in 
$\protect\theta$ between modulation pulses. These parameters were tuned such
that $\protect\theta_0$ changes by less than $5{\mkern-2mu\times\mkern-2mu}
10^{-5}$ rad between up and down pulses after 1000s.}
\label{table:dressingpulse}
\end{table}

\begin{figure}[h]
\begin{center}
\includegraphics[width=0.6\textwidth]{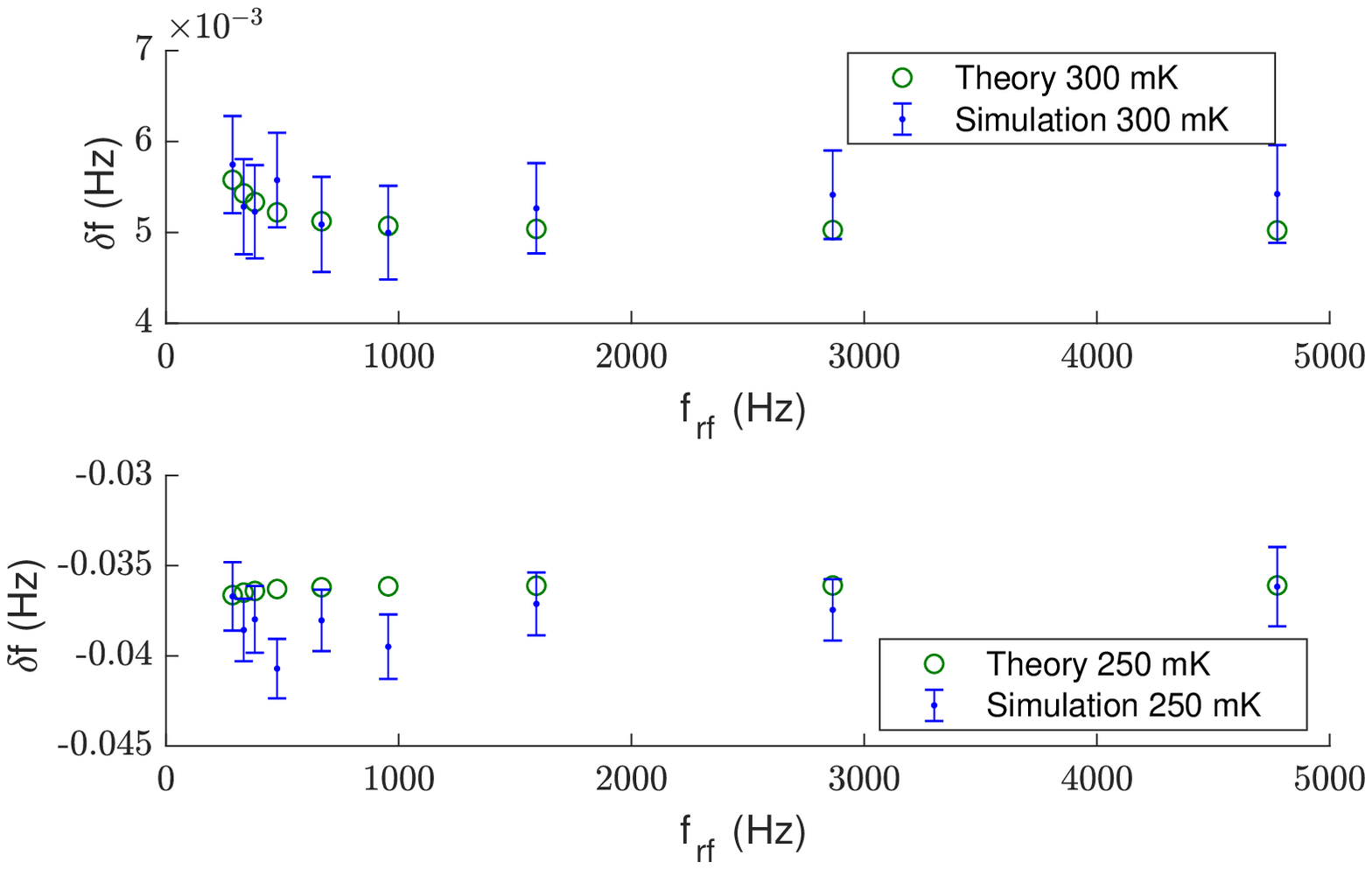}\\%

\end{center}
\caption{Linear in E frequency shift for modulated spin dressing in a rectangular cell of $%
0.4~\times~0.102~\times~0.076$~	
m (x,y,z) dimensions respectively. The spins
contained in the cell are under the influence of a modulated dressed field. The upper plot is modeling $^3$He in a superfluid helium-II bath at 250~mK in a 2~$%
\protect\mu$T holding field along the $z$ direction. The lower plot is for
the same system 300~mK in a 3~$\protect\mu$T holding field. The E field is
750~kV/cm along the $z$ direction and the magnetic field gradient is $%
2~\mu$T/m along the $x$ direction, this gradient is large compared to gradients achieved in laboratory fields in order to increase the effect above the resolution of the simulation. In this simulation the EDM is set to zero.}
\label{fig:modpulse}
\end{figure}

Pulsed modulation corrects for slow electronic drifts in the dressing
parameters, specifically drifts slower than the frequency of modulation  $f_{%
\text{fm}}\approx 1$ Hz. This is discussed further in appendix \ref{sec:APMOD}. However, modulation does not
decrease sensitivity to phase shifts compared to a dressed system that is not modulated. This includes the frequency shift due to an
electric dipole moment and any geometric frequency shifts, nor will it
increase $T_{2}$. Therefore we do not expect the analysis of the theory to
change other than by the amount prescribed by the theory due to the changing
dressing pulse that enables the modulation. The geometric phase was
simulated under the modulation parameters shown in table \ref{table:dressingpulse} and compared to the
theory, the results are shown in figure \ref{fig:modpulse}.

\bigskip 

\section{Conclusion}

An analysis of critical dressing was completed in simulations, values that
optimize the critical dressing are proposed, and are found to be in very
close agreement with theory. Furthermore an example of modulation was
proposed for a specific pulse shape over various critical dressing
frequencies and shown to be stable for periods long compared to relaxation
times in laboratories. The effect of field inhomogeneities of a dressed
system was investigated, simulations are compared to a new analytical model.

A model that incorporates Redfield-like gradient relaxation and frequency
shifts into a spin dressed system has been proposed and compared to
simulations. Good agreement between the analytical model and the simulations
is observed, allowing confidence in fast estimations without the use of a
spin dressing simulation. This provides a means of predicting observables,
in situ, that is to say, in a time frame that is much faster than the
transverse relaxation of a typical run of an experiment. Full and accurate
simulations of dressed spin systems typically take several days to complete
due to the computational complexity, this complexity is further discussed in
reference \cite{riccardoThesis}.

We highlight the prediction to cancel DC gradient relaxation by an AC gradient
relaxation, a rare scenario where two wrongs make a right. It also
gives rise to the unfortunate ability to create a relaxation rate that is faster than
the sum of the two individual rates. Nonetheless this technique holds
promise of achieving better than previously expected coherence times given
an ambient field inhomogeneity.

Finally, the model elucidates the correct approach for calculating
systematic frequency shifts; where we find the correlation function is evaluated at the dressed energy splitting, and not the intrinsic Zeeman splitting that arises from the holding field alone. Furthermore, it is shown that there is no linear in $E$ frequency shift generated by gradients in the dressing field.
\section*{Acknowledgements}
C.M. Swank would like to thank Robert Golub and for his discussions and insight into the problem. This work was supported by the National Science Foundation grants NSF-1506459 and NSF-1812340.

\bigskip

\bigskip
\appendix
\section{APPENDIX \label{sec:sigelements}}
 Here we evaluate the matrix elements given the Hamiltonian in equation \ref{eq:H}. The perturbative terms can be classified by those proportional to $\sigma
_{x},\sigma _{y},~\sigma _{z},$ $\sigma _{x}(a+a^{\dagger }),$ etc. \ For
simplicity we consider the contributions from the operators separately,
first we consider the matrix elements of $\left\langle m^{\prime }|\sigma
_{y}|m\right\rangle ,$

\begin{align}
\overline{\left\langle n^{\prime },m^{\prime }\right\vert }\sigma _{y}%
\overline{\left\vert n,m\right\rangle } &=\frac{1}{2}e^{...}a_{n^{\prime
}}^{\ast }a_{n}\left( \overline{\left\langle n_{+}^{\prime }\right\vert }%
\left\langle +\right\vert _{x}-im^{\prime }\overline{\left\langle
n_{-}^{\prime }\right\vert }\left\langle -\right\vert _{x}\right) \left(
\sigma _{y}\right) \left( \overline{\left\vert n_{-}\right\rangle }%
\left\vert +\right\rangle _{x}+im\overline{\left\vert n_{-}\right\rangle }%
_{+}\left\vert -\right\rangle _{x}\right) , \\
\overline{\left\langle n^{\prime },m^{\prime }\right\vert }\sigma _{y}%
\overline{\left\vert n,m\right\rangle } &=\frac{1}{2}e^{...}a_{n^{\prime
}}^{\ast }a_{n}\left( \overline{\left\langle n_{+}^{\prime }\right\vert }%
\left\langle +\right\vert _{x}-im^{\prime }\overline{\left\langle
n_{-}^{\prime }\right\vert }\left\langle -\right\vert _{x}\right) \left(
\left\vert +\right\rangle \left\langle -\right\vert +\left\vert
-\right\rangle \left\langle +\right\vert \right) \left( \overline{\left\vert
n\right\rangle }_{+}\left\vert +\right\rangle _{x}+im\overline{\left\vert
n\right\rangle }_{-}\left\vert -\right\rangle _{x}\right) ,  \notag \\
\overline{\left\langle n^{\prime },m^{\prime }\right\vert }\sigma _{y}%
\overline{\left\vert n,m\right\rangle } &=\frac{1}{2}e^{...}a_{n^{\prime
}}^{\ast }a_{n}\left( \overline{\left\langle n_{+}^{\prime }\right\vert }%
\left\langle +\right\vert _{x}\left\vert +\right\rangle \left\langle
-\right\vert im\overline{\left\vert n_{-}\right\rangle }\left\vert
-\right\rangle _{x}-im^{\prime }\overline{\left\langle n_{-}^{\prime
}\right\vert }\left\langle -\right\vert _{x}\left\vert -\right\rangle
\left\langle +\right\vert \overline{\left\vert n_{+}\right\rangle }%
\left\vert +\right\rangle _{x}\right) , \\
\overline{\left\langle n^{\prime },m^{\prime }\right\vert }\sigma _{y}%
\overline{\left\vert n,m\right\rangle } &=\frac{1}{2}e^{...}ia_{n^{\prime
}}^{\ast }a_{n}\left( m\overline{\left\langle n_{+}^{\prime
}|n_{-}\right\rangle }-m^{\prime }\overline{\left\langle n_{-}^{\prime
}|n_{+}\right\rangle }\right) , \\
\overline{\left\langle n^{\prime },m^{\prime }\right\vert }\sigma _{y}%
\overline{\left\vert n,m\right\rangle } &=\frac{1}{2}e^{i\left( n-n^{\prime
}\right) \omega t+i\frac{1}{2}\left( m-m^{\prime }\right) \omega
_{0}^{\prime }t}ia_{n^{\prime }}^{\ast }a_{n}\left( mJ_{n^{\prime }-n}\left( 
\frac{\gamma B_{1}}{\omega }\right) -m^{\prime }J_{n-n^{\prime }}\left( 
\frac{\gamma B_{1}}{\omega }\right) \right) .
\end{align}

Now we set $n^{\prime }=n+q$\bigskip , and sum over $n,$ giving us $%
\left\langle m^{\prime }\right\vert \sigma _{y}\left\vert m\right\rangle ,$

\begin{align}
\left\langle m^{\prime }\right\vert \sigma _{y}\left\vert m\right\rangle
=\sum_{n}\sum_{q}\frac{1}{2}e^{i\left( n-n^{\prime }\right) \omega t+i\frac{1%
}{2}\left( m-m^{\prime }\right) \omega _{0}^{\prime }t}ia_{n+q}^{\ast
}a_{n}\left( mJ_{q}\left( \frac{\gamma B_{1}}{\omega }\right) -m^{\prime
}J_{-q}\left( \frac{\gamma B_{1}}{\omega }\right) \right).
\end{align}
We keep all the terms, for now, but note here large values of $q$ do
not contribute due to the behavior of the $a_{n}$ coefficients, and $q$
determines the harmonic of the term while large harmonics will not
contribute at our desired sensitivity. We continue to simplify the expression, 
\begin{align}
\left\langle m^{\prime }\right\vert \sigma _{y}\left\vert m\right\rangle
&=\sum_{q=-\infty }^{\infty }\frac{1}{2}e^{iq\omega t+i\frac{1}{2}\left(
m-m^{\prime }\right) \omega _{0}^{\prime }t}i\left( mJ_{q}\left( \frac{%
\gamma B_{1}}{\omega }\right) -m^{\prime }J_{-q}\left( \frac{\gamma B_{1}}{%
\omega }\right) \right),
\end{align}
We note here that large values of $q$ do not contribute due to the behavior of the $a_{n}$ coefficients, and q determines the harmonic of the term. We find that it is enough to only consider the first harmonic where $q=\pm 1,$ 
\begin{align}
\left\langle m^{\prime }\right\vert \sigma _{y}\left\vert m\right\rangle &=\frac{i}{2}e^{i\frac{1}{2}\left( m-m^{\prime }\right) \omega _{0}^{\prime }t}\left( m-m^{\prime }\right) J_{0}\left( \frac{\gamma B_{1}}{\omega }\right) +\left( m+m'\right) \sin (\omega t)J_{1}\left( \frac{\gamma B_{1}}{\omega }\right).
\end{align}
 Now we turn to $\sigma _{x},$
\begin{align}
&\overline{\left\langle n^{\prime },m^{\prime }\right\vert }\sigma _{x}%
\overline{\left\vert n,m\right\rangle }, \\
\; &=\frac{1}{2}e^{i\left( n-n^{\prime }\right) \omega t+i\frac{1}{2}\left(
m-m^{\prime }\right) \omega _{0}^{\prime }t}a_{n^{\prime }}^{\ast
}a_{n}\left( \overline{\left\langle n^{\prime }\right\vert _{+}}\left\langle
+\right\vert _{x}-im\overline{\left\langle n^{\prime }\right\vert }%
_{-}\left\langle -\right\vert _{x}\right) \left( \sigma _{x}\right) \left( 
\overline{\left\vert n\right\rangle }_{+}\left\vert +\right\rangle _{x}+im%
\overline{\left\vert n\right\rangle }_{-}\left\vert -\right\rangle
_{x}\right),  \notag \\
&=\frac{1}{2}e^{...}a_{n^{\prime }}^{\ast }a_{n}\left( \overline{%
\left\langle n^{\prime }\right\vert _{+}}\left\langle +\right\vert _{x}-im%
\overline{\left\langle n^{\prime }\right\vert }_{-}\overline{\left\langle
-\right\vert _{x}}\right) \left( \left\vert +\right\rangle \left\langle
+\right\vert -\left\vert -\right\rangle \left\langle -\right\vert \right)
\left( \overline{\left\vert n\right\rangle }_{+}\overline{\left\vert
+\right\rangle }_{x}+im\overline{\left\vert n\right\rangle }_{-}\overline{%
\left\vert -\right\rangle }_{x}\right), \\
&=\frac{1}{2}e^{i\left( n-n^{\prime }\right) \omega t+i\frac{1}{2}\left(
m-m^{\prime }\right) \omega _{0}^{\prime }t}a_{n^{\prime }}^{\ast
}a_{n}\left( \overline{\left\langle n^{\prime }\right\vert _{+}}\left\langle
+\right\vert _{x}\left\vert +\right\rangle \left\langle +\right\vert 
\overline{\left\vert n\right\rangle }_{+}\overline{\left\vert +\right\rangle 
}_{x}+i^{2}m^{\prime }m\overline{\left\langle n^{\prime }\right\vert }_{-}%
\overline{\left\langle -\right\vert _{x}}\overline{\left\vert n\right\rangle 
}_{+}\overline{\left\vert -\right\rangle }_{x}\right), \\
&=\frac{1}{2}\left( 1-mm^{\prime }\right) e^{i\frac{1}{2}\left( m-m^{\prime
}\right) \omega _{0}^{\prime }t}a_{n^{\prime }}^{\ast }a_{n}\delta
_{n^{\prime }n},
\end{align}

summing over $n$ and $n^{\prime }~$we have, 
\begin{align}
\left\langle m^{\prime }\right\vert \sigma _{x}\left\vert m\right\rangle
&=\sum_{n}\frac{1}{2}\left( 1-mm^{\prime }\right) e^{i\frac{1}{2}\left(
m-m^{\prime }\right) \omega _{0}^{\prime }t}a_{n^{\prime }}^{\ast
}a_{n}\delta _{n^{\prime }n}, \\
\left\langle m^{\prime }\right\vert \sigma _{x}\left\vert m\right\rangle &=%
\frac{1}{2}\left( 1-mm^{\prime }\right) e^{i\frac{1}{2}\left( m-m^{\prime
}\right) \omega _{0}^{\prime }t}.
\end{align}

Now that we have a simple expression of the matrix elements required for $%
\left\langle \sigma _{x}\right\rangle $~we consider $\sigma _{x}\left(
a+a^{\dagger }\right) $, where we must commute $%
~a+a^{\dagger }$ with $e^{\mp \frac{1}{2}\frac{\eta }{\omega }(a^{\dagger
}-a)}.$

\begin{align}
&\sum_{m=\pm 1}\sum_{m^{\prime }=\pm 1}\overline{\left\langle n^{\prime
},m^{\prime }\right\vert }\sigma _{x}\left( a+a^{\dagger }\right) \overline{%
\left\vert n,m\right\rangle }, \\
&=\sum_{m=\pm 1}\sum_{m^{\prime }=\pm 1}\frac{1}{2}e^{i\frac{1}{2}\left(
m-m^{\prime }\right) \omega _{0}^{\prime }t}\left( \overline{\left\langle
n_{+}^{\prime }\right\vert }{\left\langle +\right\vert }_{x}-im^{\prime }%
\overline{\left\langle n_{-}^{\prime }\right\vert }{\left\langle
-\right\vert _{x}}\right) \left( \sigma _{x}\right) \left( a+a^{\dagger
}\right) \left( \overline{\left\vert n_{+}\right\rangle }{\left\vert
+\right\rangle }_{x}+im\overline{\left\vert n_{-}\right\rangle }{\left\vert
-\right\rangle }_{x}\right), \\
&=\sum_{m=\pm 1}\sum_{m^{\prime }=\pm 1}\frac{1}{2}e^{...}\left( \overline{%
\left\langle n_{+}^{\prime }\right\vert }{\left\langle +\right\vert }%
_{x}-im^{\prime }\overline{\left\langle n_{-}^{\prime }\right\vert }{%
\left\langle -\right\vert _{x}}\right) \left( \left\vert +\right\rangle
\left\langle +\right\vert -\left\vert -\right\rangle \left\langle
-\right\vert \right) \left( a+a^{\dagger }\right) \left( \overline{%
\left\vert n_{+}\right\rangle }{\left\vert +\right\rangle }_{x}+im\overline{%
\left\vert n_{-}\right\rangle }{\left\vert -\right\rangle }_{x}\right), 
\notag\\
&=\sum_{m=\pm 1}\sum_{m^{\prime }=\pm 1}\frac{1}{2}e^{i\frac{1}{2}\left(
m-m^{\prime }\right) \omega _{0}^{\prime }t}\left( \overline{\left\langle
n_{+}^{\prime }\right\vert }{\left\langle +\right\vert }_{x}\left\vert
+\right\rangle \left\langle +\right\vert \left( a+a^{\dagger }\right) 
\overline{\left\vert n_{+}\right\rangle }{\left\vert +\right\rangle }%
_{x}+i^{2}m^{\prime }m\overline{\left\langle n_{-}^{\prime }\right\vert }{%
\left\langle -\right\vert _{x}}\left( a+a^{\dagger }\right) \overline{%
\left\vert n_{-}\right\rangle }{\left\vert -\right\rangle }_{x}\right), 
\notag \\
&=\sum_{m=\pm 1}\sum_{m^{\prime }=\pm 1}e^{i\frac{1}{2}\left( m-m^{\prime
}\right) \omega _{0}^{\prime }t}\frac{1}{2}\left( \left\langle \overline{%
n_{+}^{\prime }|}\left( a+a^{\dagger }\right) \overline{|n_{+}}\right\rangle
-m^{\prime }m\left\langle \overline{n_{-}^{\prime }|}\left( a+a^{\dagger
}\right) \overline{|n_{-}}\right\rangle \right) ,
\end{align}

\bigskip Concentrating on the evaluaton of $\left( a+a^{\dagger }\right) 
\overline{\left\vert n_{\pm }\right\rangle },$ we start by writing $%
\overline{\left\vert n_{\pm }\right\rangle }$ in terms of $\left\vert
n\right\rangle ,$

\begin{align}
\left( a+a^{\dagger }\right) \overline{\left\vert n_{\pm }\right\rangle }%
=\left( a+a^{\dagger }\right) e^{\mp \frac{1}{2}\frac{\eta }{\omega }%
(a^{\dagger }-a)}\left\vert n\right\rangle,
\end{align}

for simplicity set $w=\frac{1}{2}\frac{\eta }{\omega }$, and $c=a+a^{\dagger
},$and $b=a^{\dagger }-a.$ Notice the similarities of $b$ and $c$ with the
position and momentum operators of a simple harmonic oscillator,

\begin{align}
x=\sqrt{\frac{\hbar }{2m\omega }}\left( a+a^{\dagger }\right) =\sqrt{\frac{%
\hbar }{2m\omega }}c,
\end{align}
and,
\begin{align}
p=i\sqrt{\frac{m\omega \hbar }{2}}\left( a^{\dagger }-a\right) =i\sqrt{\frac{%
m\omega \hbar }{2}}b.
\end{align}%
We know that $[x,p]=i\hbar ,$ thus,
\begin{align}
\lbrack c,b]=2.
\end{align}%
with the commutation relation
\begin{align}
\lbrack c,e^{\mp wb}]=\mp 2we^{\mp wb},
\end{align}%
the matrix elements are,
\begin{align}
\left\langle \overline{n_{\pm }^{\prime }|}\left( a+a^{\dagger }\right) 
\overline{|n_{\pm }}\right\rangle &=\left\langle n^{\prime }|\mp
2w+c|n\right\rangle , \notag \\
\left\langle \overline{n_{\pm }^{\prime }|}\left( a+a^{\dagger }\right) 
\overline{|n_{\pm }}\right\rangle &=\mp \frac{\eta }{\omega }\delta
_{n^{\prime }n}+\sqrt{n}\left\langle n^{\prime }|n-1\right\rangle +\sqrt{n+1}%
\left\langle n^{\prime }|n+1\right\rangle.
\end{align}%
Evaluating $\left\langle m^{\prime }\right\vert \sigma _{x}\left(
a+a^{\dagger }\right) \left\vert m\right\rangle $ we have
\begin{align}
\left\langle m^{\prime }\right\vert \sigma _{x}\left(
a+a^{\dagger }\right) \left\vert m\right\rangle &=\frac{1}{2}%
\sum_{n}\sum_{n^{\prime }}e^{i\left( n-n^{\prime }\right) \omega t+i\frac{1}{%
2}\left( m-m^{\prime }\right) \omega _{0}^{\prime }t}a_{n^{\prime
}}a_{n}\left( \left\langle \overline{n_{+}^{\prime }|}\left( a+a^{\dagger }\right) 
\overline{|n_{+}}\right\rangle \right.  \notag  \\ 
&\left.-m^{\prime }m\left\langle \overline{n_{-}^{\prime }|}\left( a+a^{\dagger
}\right) \overline{|n_{-}}\right\rangle \right) , \\
\left\langle m^{\prime }\right\vert \sigma _{x}\left(
a+a^{\dagger }\right) \left\vert m\right\rangle &=\frac{1}{2}\sum_{n}\sum_{n^{\prime }}e^{...}a_{n^{\prime }}a_{n}\left[ \sqrt{n}\delta _{n^{\prime }n-1}+\sqrt{n+1}\delta _{n^{\prime }n+1}-\frac{ \eta }{\omega }\delta _{n^{\prime }n} \right. \notag \\ 
&\left.-m^{\prime }m\left( \sqrt{n}\delta _{n^{\prime }n-1}+\sqrt{n+1}\delta
_{n^{\prime }n+1}+\frac{\eta }{\omega }\delta _{n^{\prime }n}\right)\right], \\
\left\langle m^{\prime }\right\vert \sigma _{x}\left( a+a^{\dagger }\right)
\left\vert m\right\rangle &=\frac{1}{2}\sum_{n}\sum_{n^{\prime
}}e^{...}a_{n^{\prime }}a_{n}\left[\sqrt{n}\delta _{n^{\prime }n-1}+\sqrt{n+1}\delta _{n^{\prime }n+1}-\frac{%
\eta }{\omega }\delta _{n^{\prime }n} \right. \notag \\ 
&\left.-m^{\prime }m\left( \sqrt{n}\delta _{n^{\prime }n-1}+\sqrt{n+1}\delta
_{n^{\prime }n+1}+\frac{\eta }{\omega }\delta _{n^{\prime }n}\right)\right].
\end{align}%
We proceed by summing over $n$,
\begin{align}
\left\langle m^{\prime }\right\vert \sigma _{x}\left( a+a^{\dagger }\right)
\left\vert m\right\rangle &=\frac{1}{2}\sum_{n}\sum_{n^{\prime }}e^{i\left( n-n^{\prime }\right)
\omega t}a_{n^{\prime }}a_{n}e^{i\frac{1}{2}\left( m-m^{\prime }\right)
\omega _{0}^{\prime }t}\\
&\times\left[ \left( \sqrt{n+1}\delta _{n^{\prime }n+1}+\sqrt{n}\delta _{n^{\prime }n-1}-\frac{\eta }{\omega }\delta _{n^{\prime}n}\left\vert m-m^{\prime }\right\vert \right) (1-m^{\prime }m)\right],\\
\left\langle m^{\prime }\right\vert \sigma _{x}\left( a+a^{\dagger }\right)
\left\vert m\right\rangle&=\frac{1}{2}e^{i\frac{1}{2}\left( m-m^{\prime }\right) \omega _{0}^{\prime
}t}\left[\left( \sum_{n}e^{i\omega t}a_{n-1}a_{n}\sqrt{n}+\sum_{n}e^{-i\omega
t}a_{n+1}a_{n}\sqrt{n+1}\right) (1-m^{\prime }m) \right. \notag \\ 
&\left.-\sum_{n}\left\vert a_{n}\right\vert ^{2}\frac{\eta }{\omega }\delta
_{nn}\left\vert m-m^{\prime }\right\vert \right], \\
\left\langle m^{\prime }\right\vert \sigma _{x}\left( a+a^{\dagger }\right)
\left\vert m\right\rangle&=\frac{1}{2}e^{i\frac{1}{2}\left( m-m^{\prime }\right) \omega _{0}^{\prime
}t}\left[ \left( e^{i\omega t}\sqrt{\lambda }+e^{-i\omega t}\sqrt{\lambda }%
\right) (1-m^{\prime }m)-\frac{\eta }{\omega }\delta _{nn}\left\vert
m-m^{\prime }\right\vert \right], \\
\left\langle m^{\prime }\right\vert \sigma _{x}\left( a+a^{\dagger }\right)
\left\vert m\right\rangle&=\lambda ^{\frac{1}{2}}e^{i\frac{1}{2}\left( m-m^{\prime }\right) \omega
_{0}^{\prime }t}\left[ \cos (\omega t)(1-m^{\prime }m)-\frac{\eta }{\lambda
^{\frac{1}{2}}\omega }\left\vert m-m^{\prime }\right\vert \right] .
\end{align}%
\bigskip Where in the second to last step with large $\lambda $ the Poisson
distribution behaves like a Dirac delta function in $n,$ numerically this is found
to be an extremely good approximation. We continue by noting that the term containing $\eta $
carries a $1/\lambda$ which is the number of photons in the field, which is very large, and so the term with $\eta $ can be neglected. We have,%
\begin{align}
\left\langle m^{\prime }\right\vert \sigma _{x}\left( a+a^{\dagger }\right)
\left\vert m\right\rangle =\lambda ^{\frac{1}{2}}e^{i\frac{1}{2}\left(
m-m^{\prime }\right) \omega _{0}^{\prime }t}\cos (\omega t)(1-mm^{\prime }).
\end{align}

Now we find
\begin{align}
&\overline{\left\langle n^{\prime },m^{\prime }\right\vert }\sigma
_{y}\left( a+a^{\dagger }\right) \overline{\left\vert n,m\right\rangle } \\
&=\frac{1}{2}e^{i\left( n-n^{\prime }\right) \omega t}e^{i\frac{1}{2}\left(
m-m^{\prime }\right) \omega _{0}^{\prime }t}\left( \overline{\left\langle
n_{+}^{\prime }\right\vert }\left\langle +\right\vert -im^{\prime }\overline{%
\left\langle n_{-}^{\prime }\right\vert }\left\langle -\right\vert \right)
\left( \sigma _{x}\right) \left( a+a^{\dagger }\right) \left( \overline{%
\left\vert n_{+}\right\rangle }\left\vert +\right\rangle +im\overline{%
\left\vert n_{-}\right\rangle }\left\vert -\right\rangle \right), \\
&=\frac{1}{2}e^{...}\left( \overline{\left\langle n_{+}^{\prime }\right\vert 
}\left\langle +\right\vert -im^{\prime }\overline{\left\langle n_{-}^{\prime
}\right\vert }\left\langle -\right\vert \right) \left( \left\vert
+\right\rangle \left\langle -\right\vert +\left\vert -\right\rangle
\left\langle +\right\vert \right) \left( a+a^{\dagger }\right) \left( 
\overline{\left\vert n_{+}\right\rangle }\left\vert +\right\rangle +im%
\overline{\left\vert n_{-}\right\rangle }\left\vert -\right\rangle \right), \notag \\
&=\frac{1}{2}e^{...}\left( \overline{\left\langle n_{+}^{\prime }\right\vert 
}\left\langle +\right\vert \left\vert +\right\rangle \left\langle
-\right\vert -im^{\prime }\overline{\left\langle n_{-}^{\prime }\right\vert }%
\left\langle -\right\vert \left\vert -\right\rangle \left\langle
+\right\vert \right) \left( a+a^{\dagger }\right) \left( \overline{%
\left\vert n_{+}\right\rangle }\left\vert +\right\rangle +im\overline{%
\left\vert n_{-}\right\rangle }\left\vert -\right\rangle \right)  \notag, \\
&=\frac{1}{2}e^{...}\left( \overline{\left\langle n_{+}^{\prime }\right\vert 
}\left\langle -\right\vert -im^{\prime }\overline{\left\langle n_{-}^{\prime
}\right\vert }\left\langle +\right\vert \right) \left( a+a^{\dagger }\right)
\left( \overline{\left\vert n_{+}\right\rangle }\left\vert +\right\rangle +im%
\overline{\left\vert n_{-}\right\rangle }\left\vert -\right\rangle \right), \\
&=e^{i\left( n-n^{\prime }\right) \omega t}e^{i\frac{1}{2}\left( m-m^{\prime
}\right) \omega _{0}^{\prime }t}\frac{i}{2}\left( m\left\langle \overline{%
n_{+}^{\prime }|}\left( a+a^{\dagger }\right) \overline{|n_{-}}\right\rangle
-m^{\prime }\left\langle \overline{n_{-}^{\prime }|}\left( a+a^{\dagger
}\right) \overline{|n_{+}}\right\rangle \right) .
\end{align}

\bigskip Therefore we must find,%
\begin{align}
\overline{\left\langle n_{\mp }^{\prime }\right\vert }\left( a+a^{\dagger
}\right) \overline{\left\vert n_{\pm }\right\rangle }=\overline{\left\langle
n_{\mp }^{\prime }\right\vert }c\overline{\left\vert n_{\pm }\right\rangle }=%
\overline{\left\langle n_{\mp }^{\prime }\right\vert }ce^{\mp \frac{1}{2}%
\frac{\eta }{\omega }b}\left\vert n_{\pm }\right\rangle .
\end{align}
With the commutation relation%
\begin{align}
\lbrack c,e^{\mp wb}]=\mp 2we^{\mp wb},
\end{align}
we have,%
\begin{align}
\overline{\left\langle n_{\mp }^{\prime }\right\vert }c\overline{\left\vert
n_{\pm }\right\rangle } &=\overline{\left\langle n_{\mp }^{\prime
}\right\vert }e^{\mp \frac{1}{2}\frac{\eta }{\omega }b}c\left\vert
n\right\rangle \mp \frac{\eta }{\omega }J_{0}(x), \\
\overline{\left\langle n_{\mp }^{\prime }\right\vert }c\overline{\left\vert
n_{\pm }\right\rangle } &=\sqrt{n}\overline{\left\langle n_{\mp }^{\prime
}\right\vert }e^{\mp \frac{1}{2}\frac{\eta }{\omega }b}\left\vert
n-1\right\rangle +\sqrt{n+1}\overline{\left\langle n_{\mp }^{\prime
}\right\vert }e^{\mp \frac{1}{2}\frac{\eta }{\omega }b}\left\vert
n+1\right\rangle \mp \frac{\eta }{\omega }J_{0}(x), \\
&=\sqrt{n}J_{n^{\prime }-n+1}\left( \frac{\gamma B_{1}}{\omega }\right)
\delta _{n^{\prime }n-1}+\sqrt{n+1}J_{n^{\prime }-n-1}\left( \frac{\gamma
B_{1}}{\omega }\right) \delta _{n^{\prime }n-1}\mp \frac{\eta }{\omega }%
J_{0}(x)\delta _{n^{\prime }n}.
\end{align}
We see that the term containing $\eta $ is negligible compared to
the other two terms. Substituting this back in and summing over $n$ and $n^{\prime }$ we have,  
\begin{align}
\left\langle m^{\prime }\right\vert \sigma _{y}\left( a+a^{\dagger }\right)
\left\vert m\right\rangle &=\frac{i}{2}%
\sum_{n}\sum_{n^{\prime }}e^{i\frac{1}{2}\left( m-m^{\prime }\right) \omega
_{0}^{\prime }t}, \\
&\times a_{n^{\prime }}a_{n}\left[ me^{i\left( n-n^{\prime }\right) \omega t}\left( \sqrt{n}J_{n^{\prime}-n+1}\left( \frac{\gamma B_{1}}{\omega }\right) \delta _{n^{\prime }n-1}+\sqrt{n+1}J_{n^{\prime }-n-1}\left( \frac{\gamma B_{1}}{\omega }\right) \delta _{n^{\prime }n-1}\right) \right. \\ 
&\left.-m^{\prime }e^{i\left( n-n^{\prime }\right) \omega t}\left( \sqrt{n}%
J_{n^{\prime }-n+1}\left( \frac{\gamma B_{1}}{\omega }\right) \delta
_{n^{\prime }n-1}+\sqrt{n+1}J_{n^{\prime }-n-1}\left( \frac{\gamma B_{1}}{%
\omega }\right) \delta _{n^{\prime }n-1}\right)\right], \\
\left\langle m^{\prime }\right\vert \sigma _{y}\left( a+a^{\dagger }\right)
\left\vert m\right\rangle&=\frac{i}{2}e^{i\frac{1}{2}\left( m-m^{\prime }\right) \omega _{0}^{\prime
}t}\left[ m\left( e^{i\omega t}\sqrt{\lambda }J_{0}\left( \frac{\gamma B_{1}}{\omega }%
\right) +e^{-i\omega t}\sqrt{\lambda }J_{0}\left( \frac{\gamma B_{1}}{\omega 
}\right) \right) \right. \notag \\ 
&\left.-m^{\prime }\left( e^{i\omega t}\sqrt{\lambda }J_{0}\left( \frac{\gamma B_{1}%
}{\omega }\right) +e^{-i\omega t}\sqrt{\lambda }J_{0}\left( \frac{\gamma
B_{1}}{\omega }\right) \right) \right], \\
\left\langle m^{\prime }\right\vert \sigma _{y}\left( a+a^{\dagger }\right)
\left\vert m\right\rangle &=ie^{i\frac{1}{2}\left( m-m^{\prime }\right)
\omega _{0}^{\prime }t}\lambda ^{\frac{1}{2}}J_{0}\left( \frac{\gamma B_{1}}{%
\omega }\right) \cos \left( \omega t\right) \left( m-m^{\prime }\right).
\end{align}

\bigskip This is expected if we consider the comparison of
the\ $q=0$ elements with $\left\langle m^{\prime }|\sigma
_{x}|m\right\rangle $ to $\left\langle m^{\prime }|\sigma
_{y}|m\right\rangle .$ We now consider,
\begin{align}
\left\langle m^{\prime }\right\vert \sigma _{z}\left( a+a^{\dagger }\right)
\left\vert m\right\rangle =\sum_{n}\sum_{n^{\prime }}\overline{\left\langle
n^{\prime },m^{\prime }\right\vert }\sigma _{z}\left( a+a^{\dagger }\right) 
\overline{\left\vert n,m\right\rangle }.
\end{align}
Concentrating on individual terms of sum we have,
\begin{align}
&\overline{\left\langle n^{\prime },m^{\prime }\right\vert }\sigma
_{z}\left( a+a^{\dagger }\right) \overline{\left\vert n,m\right\rangle } \\
&=\frac{1}{2}e^{i\left( n-n^{\prime }\right) \omega t}e^{i\frac{1}{2}\left(
m-m^{\prime }\right) \omega _{0}^{\prime }t}\left( \overline{\left\langle
n_{+}^{\prime }\right\vert }\left\langle +\right\vert -im^{\prime }\overline{%
\left\langle n_{-}^{\prime }\right\vert }\left\langle -\right\vert \right)
\left( \sigma _{x}\right) \left( a+a^{\dagger }\right) \left( \overline{%
\left\vert n_{+}\right\rangle }\left\vert +\right\rangle +im\overline{%
\left\vert n_{-}\right\rangle }\left\vert -\right\rangle \right), \\
&=\frac{i}{2}e^{...}\left( \overline{\left\langle n_{+}^{\prime }\right\vert 
}\left\langle +\right\vert -im^{\prime }\overline{\left\langle n_{-}^{\prime
}\right\vert }\left\langle -\right\vert \right) \left( -\left\vert
+\right\rangle \left\langle -\right\vert +\left\vert -\right\rangle
\left\langle +\right\vert \right) \left( a+a^{\dagger }\right) \left( 
\overline{\left\vert n_{+}\right\rangle }\left\vert +\right\rangle +im%
\overline{\left\vert n_{-}\right\rangle }\left\vert -\right\rangle \right), 
\notag \\
&=\frac{i}{2}e^{...}\left( -\overline{\left\langle n_{+}^{\prime
}\right\vert }\left\langle +\right\vert \left\vert +\right\rangle
\left\langle -\right\vert -im^{\prime }\overline{\left\langle n_{-}^{\prime
}\right\vert }\left\langle -\right\vert \left\vert -\right\rangle
\left\langle +\right\vert \right) \left( a+a^{\dagger }\right) \left( 
\overline{\left\vert n_{+}\right\rangle }\left\vert +\right\rangle +im%
\overline{\left\vert n_{-}\right\rangle }\left\vert -\right\rangle \right), 
\notag \\
&=\frac{i}{2}e^{...}\left( -\overline{\left\langle n_{+}^{\prime
}\right\vert }\left\langle -\right\vert -im^{\prime }\overline{\left\langle
n_{-}^{\prime }\right\vert }\left\langle +\right\vert \right) \left(
a+a^{\dagger }\right) \left( \overline{\left\vert n_{+}\right\rangle }%
\left\vert +\right\rangle +im\overline{\left\vert n_{-}\right\rangle }%
\left\vert -\right\rangle \right), \\
&=-e^{i\left( n-n^{\prime }\right) \omega t}e^{i\frac{1}{2}\left(
m-m^{\prime }\right) \omega _{0}^{\prime }t}\frac{i^{2}}{2}\left(
m\left\langle \overline{n_{+}^{\prime }|}\left( a+a^{\dagger }\right) 
\overline{|n_{-}}\right\rangle +m^{\prime }\left\langle \overline{%
n_{-}^{\prime }|}\left( a+a^{\dagger }\right) \overline{|n_{+}}\right\rangle
\right).
\end{align}
 Thus we find,
\begin{align}
\left\langle m^{\prime }\right\vert \sigma _{z}\left( a+a^{\dagger }\right)
\left\vert m\right\rangle =e^{i\frac{1}{2}\left( m-m^{\prime }\right) \omega
_{0}^{\prime }t}\lambda ^{\frac{1}{2}}J_{0}\left( \frac{\gamma B_{1}}{\omega 
}\right) \cos \left( \omega t\right) \left( m+m^{\prime }\right).
\end{align}

Furthermore we have,

\begin{align}
\left\langle m^{\prime }\right\vert \sigma _{z}\left\vert m\right\rangle =%
\frac{1}{2}\left( m+m^{\prime }\right) J_{0}\left( \frac{\gamma B_{1}}{\omega 
}\right)e^{i\frac{1}{2}\left( m-m^{\prime
}\right) \omega _{0}^{\prime }t}-i\left( m-m^{\prime }\right) e^{i\frac{1}{2}\left( m-m^{\prime }\right) \omega _{0}^{\prime }t}J_{1}\left( \frac{\gamma B_{1}}{\omega }\right) \sin (\omega t).
\end{align}%

\subsection{Evaluation of the perturbed expectation of $\sigma_+$ \label{sec:2ndordertheory}}
Evaluation of the perturbed expectation of $\sigma_+$ generates terms to fourth order. We only include terms to second order,
first order terms average to zero (which is not precisely true for ultracold
neutrons), and terms that include $H_{ii}H_{ik}$ oscillate fast compared to
other terms, and will be ignored. The only matrix elements that are found to
contribute are 
\begin{align}
\left\langle \sigma _{+}\right\rangle & =1-\int_{0}^{t}\int_{0}^{t^{\prime
}}H_{-+}H_{+-}dt^{\prime \prime }dt^{\prime
}-\int_{0}^{t}\int_{0}^{t^{\prime }}H_{+-}^{\ast }H_{-+}^{\ast }dt^{\prime
\prime }dt^{\prime }+\int_{0}^{t}\int_{0}^{t}H_{++}^{\ast }H_{--}dt^{\prime
}dt^{\prime \prime } \\
& -\int_{0}^{t}\int_{0}^{t^{\prime }}H_{++}^{\ast }H_{++}^{\ast }dt^{\prime
\prime }dt^{\prime }-\int_{0}^{t}\int_{0}^{t^{\prime
}}H_{--}H_{--}dt^{\prime \prime }dt^{\prime }
\end{align}%
Because the functions in $H_{jk}$ are
ultimately functions of the trajectories of stationary ensembles, is valid except for the the phases, however non-stationary phases oscillate fast and will vanish. Notice that the third integral on the right hand side has a range $[0,t]$ for both $t^{\prime }$ and $t^{\prime \prime }$. This term arises from the
first order part of the wave function. We can use the symmetry of the stationary trajectory correlation function to express this term in the common form,
\begin{align}
\int_{0}^{t}\int_{0}^{t}H_{++}^{\ast }H_{--}dt^{\prime }dt^{\prime \prime
}=-2\mathrm{Re}\left( \int_{0}^{t}\int_{0}^{t^{\prime
}}H_{--}H_{--}dt^{\prime \prime }dt^{\prime }\right).
\end{align}
we substitute this into the equation for $\left\langle \sigma
_{+}\right\rangle $ and continue,
\begin{align}
\left\langle \sigma _{+}\right\rangle &=1-\int_{0}^{t}\int_{0}^{t^{\prime
}}H_{-+}H_{+-}dt^{\prime \prime }dt^{\prime
}-\int_{0}^{t}\int_{0}^{t^{\prime }}H_{+-}^{\ast }H_{-+}^{\ast }dt^{\prime
\prime }dt^{\prime } \\
&-2\mathrm{Re}\left( \int_{0}^{t}\int_{0}^{t^{\prime
}}H_{--}H_{--}dt^{\prime \prime }dt^{\prime }\right)
-\int_{0}^{t}\int_{0}^{t^{\prime }}H_{++}^{\ast }H_{++}^{\ast }dt^{\prime
\prime }dt^{\prime }-\int_{0}^{t}\int_{0}^{t^{\prime
}}H_{--}H_{--}dt^{\prime \prime }dt^{\prime }, \\
\left\langle \sigma _{+}\right\rangle &=1-2\int_{0}^{t}\int_{0}^{t^{\prime
}}H_{-+}(t^{\prime })H_{+-}(t^{\prime \prime })dt^{\prime \prime }dt^{\prime
}-4\mathrm{Re}\left( \int_{0}^{t}\int_{0}^{t^{\prime
}}H_{--}H_{--}dt^{\prime \prime }dt^{\prime }\right) , \\
\left\langle \sigma _{+}\right\rangle &=1-2t\int_{0}^{t}H_{-+}(0)H_{+-}(\tau
)d\tau -4t\mathrm{Re}\left( \int_{0}^{t}H_{--}(0)H_{--}\left( \tau \right)
d\tau \right) .
\end{align}

In the last step we used the fact that the functions are stationary and
replaced $\tau =t^{\prime \prime }-t^{\prime }.$ In the above equation terms
that contain an oscillating phase that is fast compared to the scale of the
measurement, for example it may contain $e^{i\omega _{0}(t^{\prime
}+t^{\prime \prime })}.$, are considered negligible. Interestingly, this oscillating phase ensures that
the terms that contribute are indeed stationary. Furthermore, from the
derivation of the matrix elements found in the next section of the appendix (\ref{sec:2ndorderterms}), we find that the complex conjugate is the equivalent of changing the order of the quantum
index, thus, $H_{+-}^{\ast }=H_{-+}.$

\subsection{Evaluation of the 2nd order Matrix Elements \label{sec:2ndorderterms}}
Now we concentrate on the $\omega _{1j}(t^{\prime })\omega _{1k}(t^{\prime
\prime })$ terms,
\begin{align}
\left[ H_{+-}H_{-+}\right] _{_{\omega _{1j}\omega
_{1k}}}=\int_{0}^{t}\int_{0}^{t^{\prime }}\frac{1}{4}\omega _{1j}(t^{\prime
})\omega _{1k}(t^{\prime \prime })e^{i\omega _{0}^{\prime }t^{\prime }}\cos
(\omega t^{\prime })e^{-i\omega _{0}^{\prime }t^{\prime \prime }}\cos
(\omega t^{\prime \prime })dt^{\prime \prime }dt^{\prime }.
\end{align}

The first term can be separated into harmonics in $t^{\prime \prime }$ and $%
t^{\prime }$ for which we have,

\begin{align}
\left[ H_{+-}H_{-+}\right] _{\omega _{1j}\omega _{1k}}=&\frac{1}{4}\int \int dt^{\prime }dt^{\prime \prime }\omega _{1j}(t^{\prime
})\omega _{1k}(t^{\prime \prime })e^{i\omega _{0}^{\prime }t^{\prime }}\cos
(\omega t^{\prime })e^{-i\omega _{0}^{\prime }t^{\prime \prime }}\cos
(\omega t^{\prime \prime }), \\
\left[ H_{+-}H_{-+}\right] _{\omega _{1j}\omega _{1k}}=&\frac{1}{16}\int dt^{\prime }\int dt^{\prime \prime }\omega
_{1j}(t^{\prime })\omega _{1k}(t^{\prime \prime })\left( e^{i\omega
_{0}^{\prime }t^{\prime }+i\omega t^{\prime }}+e^{i\omega _{0}^{\prime
}t^{\prime }-i\omega t^{\prime }}\right) \left( e^{-i\omega _{0}^{\prime
}t^{\prime \prime }+i\omega t^{\prime \prime }}+e^{-i\omega _{0}^{\prime
}t^{\prime \prime }-i\omega t^{\prime \prime }}\right),  \notag \\
\left[ H_{+-}H_{-+}\right] _{\omega _{1j}\omega _{1k}}=&\frac{1}{16}\int dt^{\prime }\int dt^{\prime \prime }\omega
_{1j}(t^{\prime })\omega _{1k}(t^{\prime \prime })\left( e^{i\left( \omega _{0}^{\prime }+\omega \right) t^{\prime }-i(\omega
_{0}^{\prime }-\omega )t^{\prime \prime }}+e^{i\left( \omega _{0}^{\prime
}-\omega \right) t^{\prime }-i(\omega _{0}^{\prime }+\omega )t^{\prime
\prime }} \right. \notag \\ 
&\left.+e^{i(\omega _{0}^{\prime }+\omega )(t^{\prime }-t^{\prime \prime
})}+e^{i(\omega _{0}^{\prime }-\omega )(t^{\prime }-t^{\prime \prime })}\right), \\
\left[ H_{+-}H_{-+}\right] _{\omega _{1j}\omega _{1k}}=&\frac{1}{16}\int dt^{\prime }\int dt^{\prime \prime }\omega
_{1j}(t^{\prime })\omega _{1k}(t^{\prime \prime })\left(e^{i\omega _{0}^{\prime }(t^{\prime }-t^{\prime \prime })+i\omega (t^{\prime
}+t^{\prime \prime })}+e^{i\omega _{0}^{\prime }(t^{\prime }-t^{\prime
\prime })-i\omega (t^{\prime }+t^{\prime \prime })} \right. \notag \\ 
&\left.+e^{i(\omega _{0}^{\prime }+\omega )(t^{\prime }-t^{\prime \prime
})}+e^{i(\omega _{0}^{\prime }-\omega )(t^{\prime }-t^{\prime \prime })}\right),
\end{align}
any exponential argument containing $\omega (t^{\prime }+t^{\prime \prime
}), $ will oscillate much faster than ones with $\omega (t^{\prime
}-t^{\prime \prime }),$ and can be ignored. Furthermore we will substitute $t^{\prime \prime }-t^{\prime }=\tau,$ 
\begin{align}
\bigskip \left[ H_{+-}H_{-+}\right] _{\omega _{1j}\omega _{1k}} &=\frac{1}{16%
}\int dt^{\prime }\int dt^{\prime \prime }\omega _{1j}(t^{\prime })\omega
_{1k}(t^{\prime \prime })\left( e^{i(\omega _{0}^{\prime }+\omega
)(t^{\prime }-t^{\prime \prime })}+e^{i(\omega _{0}^{\prime }-\omega
)(t^{\prime }-t^{\prime \prime })}\right), \\
\bigskip \left[ H_{+-}H_{-+}\right] _{\omega _{1j}\omega _{1k}} &=\frac{%
\gamma ^{2}}{8}G_{x}^{2}t\int \omega _{1j}(0)\omega _{1k}(\tau )\left(
e^{-i\omega _{0}^{\prime }\tau }\cos (\omega \tau )\right) d\tau.
\end{align}

More Generally we can write, $\bigskip $%
\begin{align}
\left[ H_{+-}H_{-+}\right] _{\omega _{1j}\omega _{1k}}=\frac{\gamma ^{2}}{8}%
t\int d\tau \omega _{1j}(0)\omega _{1k}(\tau ))\left[ e^{-i\omega
_{0}^{\prime }\tau }\cos (\omega \tau )\right].
\end{align}

Here we show that the cross terms between $\omega _{0}$ and $\omega _{1}~$ for the off-diagonal elements can be neglected.

\begin{align}
\left[ H_{-+}H_{+-}\right] _{\omega _{1}^{\prime }\omega _{0}^{\prime }} &=%
\frac{1}{4}\int \int \omega _{1x}^{^{\prime }}(t^{\prime \prime })\omega
_{0x}^{\prime }(t^{\prime })\cos (\omega t^{\prime \prime })e^{i\omega
_{0}^{\prime }(t^{\prime \prime }-t^{\prime })}dt^{\prime }dt^{\prime \prime
} \\
&+\frac{1}{4}\int \int \omega _{1x}^{^{\prime }}(t^{\prime })\omega
_{0x}^{\prime }(t^{\prime \prime })\cos (\omega t^{\prime })e^{i\omega
_{0}^{\prime }(t^{\prime \prime }-t^{\prime })}dt^{\prime }dt^{\prime \prime
}, \\
\left[ H_{-+}H_{+-}\right] _{\omega _{1}^{\prime }\omega _{0}^{\prime }}&=\frac{1}{8}\int \int \omega _{1x}^{^{\prime }}(t^{\prime \prime })\omega
_{0x}^{\prime }(t^{\prime })(e^{i\omega t^{\prime \prime }}+e^{-i\omega
t^{\prime \prime }})e^{i\omega _{0}^{\prime }(t^{\prime \prime }-t^{\prime
})}dt^{\prime }dt^{\prime \prime } \\
&+\frac{1}{8}\int \int \omega _{1x}^{^{\prime }}(t^{\prime })\omega
_{0x}^{\prime }(t^{\prime \prime })(e^{i\omega t^{\prime \prime
}}+e^{-i\omega t^{\prime \prime }})e^{i\omega _{0}^{\prime }(t^{\prime
\prime }-t^{\prime })}dt^{\prime }dt^{\prime \prime }.
\end{align}
We set $\tau =t^{\prime \prime }-t^{\prime }$, 
\begin{align}
\left[ H_{-+}H_{+-}\right] _{\omega _{1}^{\prime }\omega _{0}^{\prime }} &=%
\frac{1}{8}\int \int \omega _{1x}^{^{\prime }}(\tau +t^{\prime })\omega
_{0x}^{\prime }(t^{\prime })(e^{i\omega (\tau +t^{\prime })}+e^{-i\omega
(\tau +t^{\prime })})e^{i\omega _{0}^{\prime }\tau }dt^{\prime }dt^{\prime
\prime }, \\
&+\frac{1}{8}\int \int \omega _{1x}^{^{\prime }}(t^{\prime })\omega
_{0x}^{\prime }(\tau +t^{\prime })(e^{i\omega t^{\prime }}+e^{-i\omega
t^{\prime }})e^{i\omega \tau }dt^{\prime }dt^{\prime \prime } \\
\left[ H_{-+}H_{+-}\right] _{\omega _{1}^{\prime }\omega _{0}^{\prime }}&=\frac{1}{8}\int \int \omega _{1x}^{^{\prime }}(\tau +t^{\prime })\omega
_{0x}^{\prime }(t^{\prime })(e^{i(\omega +\omega _{0}^{\prime })(\tau
+t^{\prime })}+e^{i\omega (-\omega +\omega _{0}^{\prime })(\tau +t^{\prime
})})dt^{\prime }dt^{\prime \prime } \\
&+\frac{1}{8}\int \int \omega _{1x}^{^{\prime }}(t^{\prime })\omega
_{0x}^{\prime }(\tau +t^{\prime })(e^{i\omega t^{\prime }}+e^{-i\omega
t^{\prime }})e^{i\omega \tau }dt^{\prime }dt^{\prime \prime }.
\end{align}
We see that all terms contain oscillating phases that allow the result to be
neglected,
\begin{align}
\left[ H_{-+}H_{+-}\right] _{\omega _{1}^{\prime }\omega _{0}^{\prime
}}\approx 0.
\end{align}
Here we examine the cross frequency terms of the diagonal elements ($%
\sigma _{z}$), and show they can be neglected,
\begin{align}
\left[ H_{--}H_{--}\right] _{\omega _{1}^{\prime }\omega _{0}^{\prime }}=%
\frac{1}{4}\int \int \omega _{1z}^{^{\prime }}(t^{\prime \prime })\omega
_{0z}^{\prime }(t^{\prime })\cos (\omega t^{\prime \prime })dt^{\prime
}dt^{\prime \prime }+\frac{1}{4}\int \int \omega _{1z}^{^{\prime
}}(t^{\prime })\omega _{0z}^{\prime }(t^{\prime \prime })\cos (\omega
t^{\prime })dt^{\prime }dt^{\prime \prime },
\end{align}
substituting\bigskip\ $\tau =t^{\prime \prime }-t^{\prime }$,
\begin{align}
\left[ H_{--}H_{--}\right] _{\omega _{1}^{\prime }\omega _{0}^{\prime }}&=%
\frac{1}{4}\int \int \omega _{1z}^{^{\prime }}(\tau +t^{\prime })\omega
_{0z}^{\prime }(t^{\prime })\left( e^{i\omega (\tau +t^{\prime
})}+e^{-i\omega (\tau +t^{\prime })}\right) dt^{\prime }dt^{\prime \prime }
\\
&+\frac{1}{4}\int \int \omega _{1z}^{^{\prime }}(t^{\prime })\omega
_{0z}^{\prime }(t^{\prime \prime })\left( e^{i\omega t^{\prime
}}+e^{-i\omega t^{\prime }}\right) dt^{\prime }dt^{\prime \prime }.
\end{align}
Again, all terms contain a rapidly oscillating phase so we have,
\begin{align}
\left[ H_{--}H_{--}\right] _{\omega _{1}^{\prime }\omega _{0}^{\prime
}}\approx 0.
\end{align}
The evaluation of the matrix elements for $\omega'_0\omega'_0$ terms are not included here. However, identical derivations are found in reference \cite{redfield,GolubSteyerlRedfield,pignol2015} with the exception of the factor $J_0(x)^2$. 
Now we show evaluation of the matrix elements that contribute from the terms where $q=\pm 1$. Several of these terms cancel due to fast oscillating factors, or by symmetry, here we consider terms that contribute in principle,
\begin{align}
&\left[ H_{-+}H_{+-}\right] _{\omega _{1x}^{\prime }\omega _{0zq}^{\prime }}\\
&=\int \int \frac{i}{2}e^{i\omega _{0}^{\prime }\left( t-t^{\prime }\right) }\left[ \sin (\omega t^{\prime })\cos (\omega t)\omega _{1x}(t)\omega _{0qz}(t^{\prime })-\sin (\omega t)\cos (\omega t^{\prime })\omega _{1x}(t^{\prime })\omega _{0qz}(t)\right] dtdt^{\prime },\\
&=\int \int \frac{i}{2}e^{i\omega _{0}^{\prime }(t-t^{\prime })}\omega _{1x}(t^{\prime })\omega _{0qz}(t)\frac{i}{4}\left[ \left( e^{i\omega t^{\prime }}-e^{-i\omega t^{\prime }}\right) \left( e^{i\omega t}+e^{-i\omega t}\right)-\left( e^{i\omega t}-e^{-i\omega t}\right) \left( e^{i\omega t^{\prime }}+e^{-i\omega t^{\prime }}\right) \right] dtdt^{\prime },\\
&=\int \int \frac{i}{2}e^{i\omega _{0}^{\prime }(t-t^{\prime })}\frac{i}{4}\omega _{1x}(t^{\prime })\omega _{0qz}(t)\left[ \left( e^{i\omega t^{\prime }-i\omega t}-e^{i\omega t-i\omega t^{\prime }}\right) -\left( e^{i\omega t-i\omega t^{\prime }}-e^{i\omega t^{\prime }-i\omega t}\right) \right] dtdt^{\prime },\\
&=-t\int \frac{i}{2}e^{-i\omega _{0}^{\prime }\tau }\sin (\omega \tau )\omega _{1x}(0)\omega _{0zq}(\tau )d\tau.\\
\end{align}
We point out that this is a phase and freuquency shifted sine transform. This translates into the real part of the difference in the spectrum shifted by $\omega\pm\omega'_0$ and when $\omega>>\omega'_0$ these terms are negligible. A similar derivation is completed for the squared terms of the off-diagonal components,
\begin{align}
&\left[ H_{-+}H_{+-}\right] _{\omega _{0zq}^{\prime }\omega _{0zq}^{\prime }}\\
&=\int \int e^{i\omega _{0}^{\prime }\left( t-t^{\prime }\right) }\left[ \sin (\omega t^{\prime })\sin (\omega t)\omega _{0zq}(t)\omega _{0zq}(t^{\prime })(t)\right] dtdt^{\prime },\\
&=-\int \int e^{i\omega _{0}^{\prime }(t-t^{\prime })}\left[ \frac{1}{4}\left( e^{i\omega t^{\prime }-i\omega t}-e^{i\omega t-i\omega t^{\prime }}\right) \left( e^{i\omega t}-e^{-i\omega t}\right) \right] dtdt^{\prime },\\
&=-\int \int e^{i\omega _{0}^{\prime }(t-t^{\prime })}\left[ \left( e^{i\omega t^{\prime }-i\omega t}-e^{i\omega t-i\omega t^{\prime }}\right) \right] dtdt^{\prime },\\
&=-\frac{it}{2}\int e^{-i\omega _{0}^{\prime }\tau }\sin (\omega \tau )\omega _{0zq}(0)\omega _{0zq}(\tau )d\tau.
\end{align}
Again, we point out that this will be highly suppressed as it goes with the difference in the spectrum at $\omega\pm\omega'_0$. Which is negligible when $\omega>>\omega'_0$.

The last term that contributes, in principle, is the diagonal components for $q=\pm 1$. The only non-vanishing term is the squared term from $\sigma_y$. It can be shown to be,
\begin{align}
&\left[ H_{--}H_{--}\right] _{\omega _{0yq}^{\prime }\omega _{0yq}^{\prime }}=-\frac{it}{2}\int \sin (\omega \tau )\omega _{0yq}(0)\omega _{0yq}(\tau )d\tau
\end{align}
From symmetry, this term is zero when we take the real part to find the relaxation.

\subsection{Simplification of the DC field diagonal contribution to $T_2$ \label{sec:T2dc}}
Starting from equation \ref{eq:genT20} we continue by expanding the exponential in a series and taking the leading order terms,
\begin{align}
\left\langle \sigma _{+}\right\rangle & =e^{i\gamma \int_{0}^{t^{\prime
}}\left( \frac{1}{\omega }J_{1}\left( \frac{\gamma \left\langle
B_{1}\right\rangle }{\omega }\right) B_{0}\omega _{1x}\left( t^{\prime
}\right) +1J_{0}\left( \frac{\gamma \left\langle B_{1}\right\rangle }{\omega 
}\right) \delta B_{0}(t^{\prime })\right) d} ,\\
\left\langle \sigma _{+}\right\rangle & \approx 1-i\gamma \int_{0}^{t^{\prime }}\left( \frac{\gamma }{\omega }%
J_{1}\left( \frac{\gamma \left\langle B_{1}\right\rangle }{\omega }\right)
B_{0}\omega _{1x}\left( t^{\prime }\right) +\gamma J_{0}\left( \frac{\gamma
\left\langle B_{1}\right\rangle }{\omega }\right) \delta B_{0}(t)\right)
dt^{\prime },\\
& -\frac{\gamma ^{2}}{2}\int_{0}^{t^{\prime }}\left( \frac{\gamma }{\omega }%
J_{1}\left( \frac{\gamma \left\langle B_{1}\right\rangle }{\omega }\right)
B_{0}\omega _{1x}\left( t^{\prime }\right) +\gamma J_{0}\left( \frac{\gamma
\left\langle B_{1}\right\rangle }{\omega }\right) B_{0}^{^{\prime
}}(t^{\prime })\right) dt^{\prime }\times ... \\
& \int_{0}^{t^{\prime }}\left( \frac{\gamma }{\omega }J_{1}\left( \frac{%
\gamma \left\langle B_{1}\right\rangle }{\omega }\right) B_{0}\omega
_{1x}\left( t^{\prime }\right) +\gamma J_{0}\left( \frac{\gamma \left\langle
B_{1}\right\rangle }{\omega }\right) B_{0}^{^{\prime }}(t^{\prime })\right)
dt^{\prime \prime }.
\end{align}%
Now we write $x=\frac{\gamma \left\langle B_{1}\right\rangle }{\omega }%
,$ to signify that there is no more time dependence within the Bessel
functions. From the definition of $\delta B_{0}(t)$ we see that the first
term averages to zero, for the last term on the right hand side we find,
\begin{align}
\left\langle \sigma _{+}\right\rangle & \approx 1-\frac{\gamma ^{2}}{2}%
J_{0}\left( x\right) ^{2}\int^{t}\delta B_{0}(t^{\prime })dt^{\prime
}\int^{t^{\prime }}\delta B_{0}(t^{\prime \prime })dt^{\prime \prime }, \\
& -\frac{J_{1}(x)^{2}\omega _{0}^{2}}{2\omega ^{2}}\int_{0}^{t}\omega
_{1x}\left( t^{\prime }\right) dt^{\prime }\int_{0}^{t^{\prime }}\omega
_{1x}\left( t^{\prime \prime }\right) dt^{\prime \prime }, \\
& +\frac{\gamma ^{2}J_{0}\left( x\right) J_{1}\left( x\right) B_{0}}{\omega }%
\int^{t}\omega _{1x}\left( t^{\prime }\right) dt^{\prime }\int^{t^{\prime
}}\delta B_{0}(t^{\prime \prime })dt^{\prime \prime }.
\end{align}
We recognize that all functions are mechanically stationary and proceed to
write them in terms of the correlation functions of the variable $\tau
=t^{\prime \prime }-t^{\prime },$
\begin{align}
\left\langle \sigma _{+}\right\rangle & \approx 1-\frac{\gamma ^{2}}{2}%
J_{0}(x)^{2}\left( t-\left\vert \tau \right\vert \right)
\int_{-t}^{t}B_{0}^{\prime }(0)B_{0}^{^{\prime }}(\tau )d\tau,\\
& -\frac{J_{1}(x)^{2}\omega _{0}^{2}}{2\omega ^{2}}\left( t-\left\vert \tau
\right\vert \right) \int_{-t}^{t}\omega _{1}^{\prime }\left( 0\right) \omega
_{1}^{\prime }\left( \tau \right) d\tau, \\
& +\frac{\gamma J_{0}\left( x\right) J_{1}\left( x\right) B_{0}}{\omega }%
\left( t-\left\vert \tau \right\vert \right) \int_{-t}^{t}\omega
_{1}^{\prime }\left( 0\right) B_{0}^{^{\prime }}(\tau )d\tau.
\end{align}
$~$Now we take the limit that $\tau _{c}$ is much smaller than $t$, but at
time $t$ the decay is small, allowing us to write the results in the usual
form of a Transform. From this simplification we can write,
\begin{align}
\left\langle \sigma _{+}\right\rangle & \approx 1-\gamma
^{2}J_{0}(x)^{2}t\int_{0}^{\infty }B_{0}^{\prime }(0)B_{0}^{^{\prime }}(\tau
)d\tau,  \\
& -\frac{J_{1}(x)^{2}\omega _{0}^{2}}{\omega ^{2}}t\int_{0}^{\infty }\omega
_{1}^{\prime }\left( 0\right) \omega _{1}^{\prime }\left( \tau \right) d\tau, 
\\
& +2\frac{\gamma ^{2}J_{0}\left( x\right) J_{1}\left( x\right) B_{0}}{\omega 
}t\int_{0}^{\infty }\omega _{1}^{\prime }\left( 0\right) B_{0}^{^{\prime
}}(\tau )d\tau.
\end{align}
This result is used to derive equation~\ref{eq:dcT2}.
\subsection{Modulation alternative}
\label{sec:APMOD}
The pulsed modulation is optimized to maximize sensitivity while correcting for
linear drifts below about $~f_{\text{fm}}=1~$Hz. However noise, either
external noise, or noise generated by field inhomogeneities can be further
reduced by cosine modulation in conjunction with a sharp band pass around $%
f_{\text{fm}}$ to 'lock-in' to the signal. It is expected that in general
the ultimate sensitivity will be larger for pulsed frequency
modulation. This is due to the strict requirement on the inhomogeneities of the
dressing field from the requirement of transverse relaxation time, simultaneously ensuring
minimal noise from spin dressing inhomogeneities.

\bibliographystyle{ieeetr}
\bibliography{SwankBibliography}

\end{document}